\documentclass[%
 jmp,
 amsmath,amssymb,
preprint,
]{revtex4-1}

\newcommand{\beqn}{\begin{eqnarray}}
\newcommand{\eeqn}{\end{eqnarray}}

\DeclareMathOperator\Pf{Pf}

\DeclareMathOperator{\Tr}{Tr}
\DeclareMathOperator{\erfc}{erfc}
\DeclareMathOperator{\erf}{erf}
\DeclareMathOperator{\diag}{diag}

\newcommand{\be}{\begin{equation}}
\newcommand{\ee}{\end{equation}}

\def\Xint#1{\mathchoice
   {\XXint\displaystyle\textstyle{#1}}%
   {\XXint\textstyle\scriptstyle{#1}}%
   {\XXint\scriptstyle\scriptscriptstyle{#1}}%
   {\XXint\scriptscriptstyle\scriptscriptstyle{#1}}%
   \!\int}
\def\XXint#1#2#3{{\setbox0=\hbox{$#1{#2#3}{\int}$}
     \vcenter{\hbox{$#2#3$}}\kern-.5\wd0}}

\def\dashint{\Xint-}

\newcommand{\nn}{\nonumber}

\newcommand{\bea}{\begin{eqnarray}}
\newcommand{\eea}{\end{eqnarray}}
\newcommand{\beq}{\begin{equation}}
\newcommand{\eeq}{\end{equation}}

\usepackage[utf8]{inputenc}
\usepackage{amsmath}
\usepackage{hyperref}
\usepackage{graphicx}
\usepackage{amsfonts}
\usepackage{amsthm}
\usepackage{cases}
\usepackage{bm}
\usepackage{amssymb}

\hypersetup{
    colorlinks=true,       
    linkcolor=red,          
    citecolor=blue,        
    filecolor=magenta,      
    urlcolor=cyan           
}

\newcommand{\moy}[1]{\ensuremath{\left\langle #1 \right\rangle}}

\usepackage{graphicx}
\usepackage{dcolumn}
\usepackage{bm}

\usepackage[utf8]{inputenc}
\usepackage[T1]{fontenc}
\usepackage{mathptmx}
\usepackage{etoolbox}

\makeatletter
\def\@email#1#2{%
 \endgroup
 \patchcmd{\titleblock@produce}
  {\frontmatter@RRAPformat}
  {\frontmatter@RRAPformat{\produce@RRAP{*#1\href{mailto:#2}{#2}}}\frontmatter@RRAPformat}
  {}{}
}%
\makeatother
\begin{document}

\preprint{AIP/123-QED}

\title[Superposition of Random Plane Waves in High Spatial Dimensions: Random Matrix Approach to Landscape Complexity]{Superposition of Random Plane Waves in High Spatial Dimensions:\\ Random Matrix Approach to Landscape Complexity}
\author{Bertrand Lacroix-A-Chez-Toine}
\thanks{These two authors contributed equally}
 \affiliation{Department of Mathematics, King’s College London, London
WC2R 2LS, United Kingdom}

\email{bertrand.lacroix\_a\_chez\_toine@kcl.ac.uk}
\author{Sirio Belga Fedeli}%
 \thanks{These two authors contributed equally}
\affiliation{%
Department of Mathematics, King’s College London, London
WC2R 2LS, United Kingdom
} 
 
\affiliation{Present address: Simons Center for Systems Biology, School of Natural Sciences, Institute for Advanced Study, Princeton, NJ 08540, U.S.A.}%
\email{fedeli@ias.edu}


\author{Yan V. Fyodorov}
\affiliation{%
Department of Mathematics, King’s College London, London
WC2R 2LS, United Kingdom
}
\affiliation{L. D. Landau Institute for Theoretical Physics, Semenova 1a, 142432 Chernogolovka, Russia}

\email{yan.fyodorov@kcl.ac.uk}

\date{\today}

\begin{abstract}
 Motivated by current interest in understanding statistical properties of random landscapes in  high-dimensional spaces, we consider a model of the landscape in   $\mathbb{R}^N$ obtained by superimposing $M>N$ plane waves of random wavevectors and amplitudes.
  For this landscape we show how to compute  the "annealed complexity" controlling the asymptotic growth rate of the mean number of stationary points as $N\to \infty$ at fixed ratio $\alpha=M/N>1$. The framework of this computation requires us to study spectral properties of $N\times N$ matrices $W=KTK^T$, where $T$ is diagonal with $M$ mean zero i.i.d. real normally distributed entries, and all $MN$ entries of $K$ are also i.i.d. real normal random variables. We suggest to call the latter Gaussian Marchenko-Pastur Ensemble, as such matrices appeared in the seminal 1967 paper by those authors.   We compute the associated mean spectral density and evaluate some moments and correlation functions involving products of  characteristic polynomials for such and related matrices.
\end{abstract}

\maketitle


\section{Introduction}

The problem of characterising the statistical properties of high-dimensional random functions, frequently called ''random landscapes'',  originated in the theory of disordered systems such as  spin glasses, see \cite{CCrev} for an accessible introduction, and gradually became popular beyond the original setting, finding numerous applications in such diverse fields as cosmology \cite{cosmol1,cosmol2}, machine learning via deep neural networks \cite{Choromanska_et_al,Bask1}, and large-size inference problems in statistics \cite{BMMN19,RABC20,MBAB20,FT21}. One of the simplest yet nontrivial landscape characteristics are the so-called {\it landscape complexities} given by the rates of exponential growth  of the number of stationary points (minima, maxima, saddles of a particular index) with the number $N$ of dimensions of the associated landscape.  Extracting those rates for various models of random spin-glass landscapes initially attracted attention in theoretical physics literature \cite{CGP98,CGG99,F04,BD07,FW07}, and
became a focus of vigorous activity in recent years, see \cite{ABAC13a,ABAC13b,FN12,S17,FLDRT18,R2020,FT20,SS21,BABM21a,GK21} and references therein. 

The fully controlled (and eventually rigorous) evaluation of complexities 
has been so far demonstrated only for a few types of landscapes whose random parts possess high level of statistical symmetry in the underlying  $N$-dimensional space. One of the most paradigmatic examples is the landscape
\be
{\cal V}({\bf x})=\frac{\mu}{2}|{\bf x}|^2+V({\bf x})\;,\,\,{\bf x}\in \mathbb{R}^N \label{landscape}
\ee
where $\mu$ is the strength of a isotropic harmonic confinement and $V({\bf x})$ is a Gaussian-distributed random function which is  translationally and rotationally invariant and characterized by the covariance structure 
\be
\moy{V({\bf x}_1)V({\bf x}_2)}=N\,\Gamma\left(\frac{({\bf x}_1-{\bf x}_2)^2}{2N}\right)\;.\label{trans_rot_inv}
\ee
In such a case the associated Hessian $W_{ij}({\bf x})=-\partial_{x_i,x_j}V({\bf x})$ turns out to be given by a matrix simply related \cite{F04,F15} to the classical Gaussian Orthogonal Ensemble of random matrices. Similar Hessians appear also in studies of rotationally-invariant random functions confined to the surface of a high-dimensional sphere \cite{ABAC13a,ABAC13b}. In both situations the exploitation of the so-called Kac-Rice formula allowsed to perform explicit computations of the so-called ''annealed'' (i.e. related to the mean number of stationary points) complexities. Moreover, one can further show that annealed and  quenched (i.e. characterizing typical realizations of the landscape) complexities coincide for certain class of potentials $V({\bf x})$ \cite{S17,SS21}, though in general they may differ.

The simplest, yet informative characteristics of the landscape is just the total annealed complexity of all possible stationary points.  In particular, studying such complexity revealed that the invariant landscapes (\ref{landscape})-(\ref{trans_rot_inv}) undergo a "topology trivialisation transition" from the glassy phase with exponentially many stationary points for small values of the ratio $\mu/\sqrt{\Gamma''(0)}<1$, to the phase where there is typically only a single minimum for $\mu/\sqrt{\Gamma''(0)}>1$. 
Similar transitions are also operative in random potentials confined to  high-dimensional spheres \cite{RABC20,F15,FLD14,BCNS21}.

The goal of this article is to suggest another class of random high-dimensional landscapes amenable to well-controlled treatment of the annealed complexity. Our construction starts with the same representation  \eqref{landscape} but proceeds with choosing the random part in the form
of a superposition of random plane waves:
\be
V({\bf x})=\sum_{a=1}^M \left[u_1^{(a)}\cos({\bf k}_a\cdot {\bf x})+u_2^{(a)}\sin({\bf k}_a\cdot {\bf x})\right]\;,
\label{mod_superpos}
\ee
where the $u_{1,2}^{(a)}$ are $M$ i.i.d. normal (mean zero, variance one) random variables and the $M$ vectors ${\bf k}_a=(k_{1,a},\cdots,k_{N,a})$ are all i.i.d., with  $N$ real independent, normally distributed (mean zero, variance $1/N$)  random components $k_{i,a},\, i=1,\ldots,N$. We henceforth denote $\alpha=M/N$.

Note that the random functions defined via \eqref{mod_superpos} are not Gaussian-distributed due to additional randomness in the choice of wavevectors ${\bf k}_a$. Note also that for
$\alpha<1$ at every realization the function $V({\bf x})$ is constant in the $M-N$ subspace orthogonal to one spanned by $M$ random wavevectors ${\bf k}_a$. To avoid this situation and to have $V({\bf x})$ varying in the whole space we therefore restrict our computation of the landscape complexity to the case $\alpha\ge 1$. It is also worth mentioning that stationary points in not unrelated random superpositions of plane waves in low spatial dimension $N=2$ is currently under intensive study as a universal model for high-energy eigenfunctions of the Laplace operator on generic compact Riemannian manifolds, see \cite{BCW2019} and references therein. Finally, for low values of $M,N$ the model \eqref{mod_superpos} describes periodic systems such as crystals in presence of quenched impurities and was much studied, mostly for its thermodynamics and dynamics, especially in the context of vortex lattices in superconductors, for reviews see e.g. \cite{BFGLV94,LD10}.\\[1ex]
We believe that the suggested landscape in high dimensions represents a new interesting type worth consideration. In particular

\begin{enumerate}
\item  such landscapes still allow an explicit analytical computation of the total annealed complexity beyond the class of universality of previously studied potentials satisfying \eqref{trans_rot_inv}.
We will see that the arising $\mu-$dependence will be quite different. In particular, the landscape will not show the sharp topology trivialization transition at any finite value of $\mu$. 
\item In contrast to the translationally-invariant random fields, which are extremely difficult to simulate even for moderately big $N\sim 10$, a superposition of random plane waves is easily constructed numerically.
\end{enumerate}

As will be demonstrated below, the mean total number of stationary points in the landscape can be expressed in terms of the random matrix average
\be
\moy{{\cal N}_{\rm tot}(\mu;\alpha)}=\frac{\moy{|\det(\mu\, \mathbb{I}-W)|}}{\mu^N}\;,\;\;\mu>0\,,\label{N_tot_abs_charpol}
\ee
where the Hessian matrix $W$ can be represented as $W=K T K^T$ and turns out to belong to the "Gaussian Marchenko-Pastur Ensemble" (GMPE) that we define as
\begin{equation}\label{GMPE_Matrix}
W_{ij}=\sum_{a=1}^M T_a k_{ia}k_{ja}\;,\;\;1\leq i,j\leq N\;,
\end{equation}
where the variables $k_{ia}$ for $1\leq i\leq N$ and $1\leq a \leq M$ (considered as the entries of matrix  $K$) are independent real normal random variables (mean zero, variance $1/N$), whereas the $M\times M$ matrix $T$ is diagonal with $N(0,1)$  normal random elements $T_a,\, a=1,\ldots,M$. Here and in the following, we denote $\moy{\cdots}_T$ the average with respect to the diagonal matrix $T$, $\moy{\cdots}_K$ the average with respect to the matrix $K$ and $\moy{\cdots}$ the average with respect to both $K$ and $T$. The suggested name for this random matrix ensemble \cite{BF21} seems appropriate to us as this type of matrices appeared already in the seminal Marchenko-Pastur paper \cite{MP67}, though does not seem to be under much study ever since.

Note that had we replaced the diagonal matrix $T$ with the identity matrix one would arrive to the standard Wishart Ensemble whose eigenvalue density at $N\to \infty$ is given by the famous Marchenko-Pastur distribution. Taking any positive-definite $T$ would provide a natural generalization of the Wishart ensemble. However the sign-indefinite nature of $T$ in our case makes the properties of the resulting matrices $W$ very different from the Wishart case. We believe the resulting ensemble GMPE is interesting in its own right and we further study  its mean eigenvalue density as $N\to \infty$ as well as correlation properties of its characteristic polynomial for both fixed and random diagonal matrix $T$. While the main focus of this article is on the landscape complexity, hence on the real symmetric matrices with the Dyson index $\beta=1$ (corresponding to $K$ being a matrix with real independent Gaussian elements), from the point of view of Random Matrix Theory (RMT) it is natural  to extend the  definition of the Gaussian Marchenko-Pastur Ensemble to the Dyson index $\beta=2$ by considering $K$ with normal complex independent elements \cite{footnote}. 
The $p$-point correlation function of the associated characteristic polynomial for a fixed diagonal matrix $T$ is then defined as 
\begin{equation}\label{charpol}
{\cal Z}_{\beta,p}({\bm \lambda};T)=\moy{\prod_{j=1}^p \det(\lambda_j\,\mathbb{I}-W)}_K\;,
\end{equation}
where we take the vector ${\bm \lambda}=(\lambda_1,\cdots,\lambda_p)$ of spectral parameters to have real elements. For random $T$, the correlation function in \eqref{charpol} is defined with  additionally taking the expectation with respect to the distribution of $T_a$. We will provide some explicit results for the correlation functions for fixed $T$ at coinciding points $\lambda_1=\ldots=\lambda_p$ for any integer $p$ and $N$, whereas for random $T$ we will restrict ourselves to $p=1,2$.  Treating in the latter case for non-coinciding points $\lambda_1 \ne \lambda_2$ we  demonstrate that the correlation structure emerging after averaging over random $T_a$ is quite different from the one known for both the standard invariant ensembles of random matrices and for ensembles with i.i.d. entries. The correlations of characteristic polynomials in the latter two types of RMT ensembles have been under intensive studies in the last two decades, see \cite{BH1,BH2,FStra,StraF,BDS,BS06,K1,K2,Shch1,Shch2,Af1,Af2}.

The rest of the paper is organised as follows. In section \ref{sec_res_sum}, we summarise the main results obtained in this paper. In section \ref{sec_count}, we consider the total annealed complexity of the landscape \eqref{landscape} with the random potential defined in \eqref{mod_superpos}. In the following sections, we characterise the Gaussian Marchenko-Pastur Ensemble associated to this counting problem by computing its mean spectral density in section \ref{sec_dens} and correlation function of its characteristic polynomial defined in \eqref{charpol} in section \ref{sec_pol}. Finally, in section \ref{sec_conclu}, we discuss and interpret our results and give some perspective on further investigation of this random landscape. Some details of the computations are relegated to Appendices \ref{app_p_1}, \ref{app_Z_2_b_N_fin} and \ref{app_LD_Z_b_2}.

\section{Main results}\label{sec_res_sum}

\subsection{Results on the counting problem}

\begin{figure}[t!]
\centering
 \includegraphics[width=.8\textwidth]{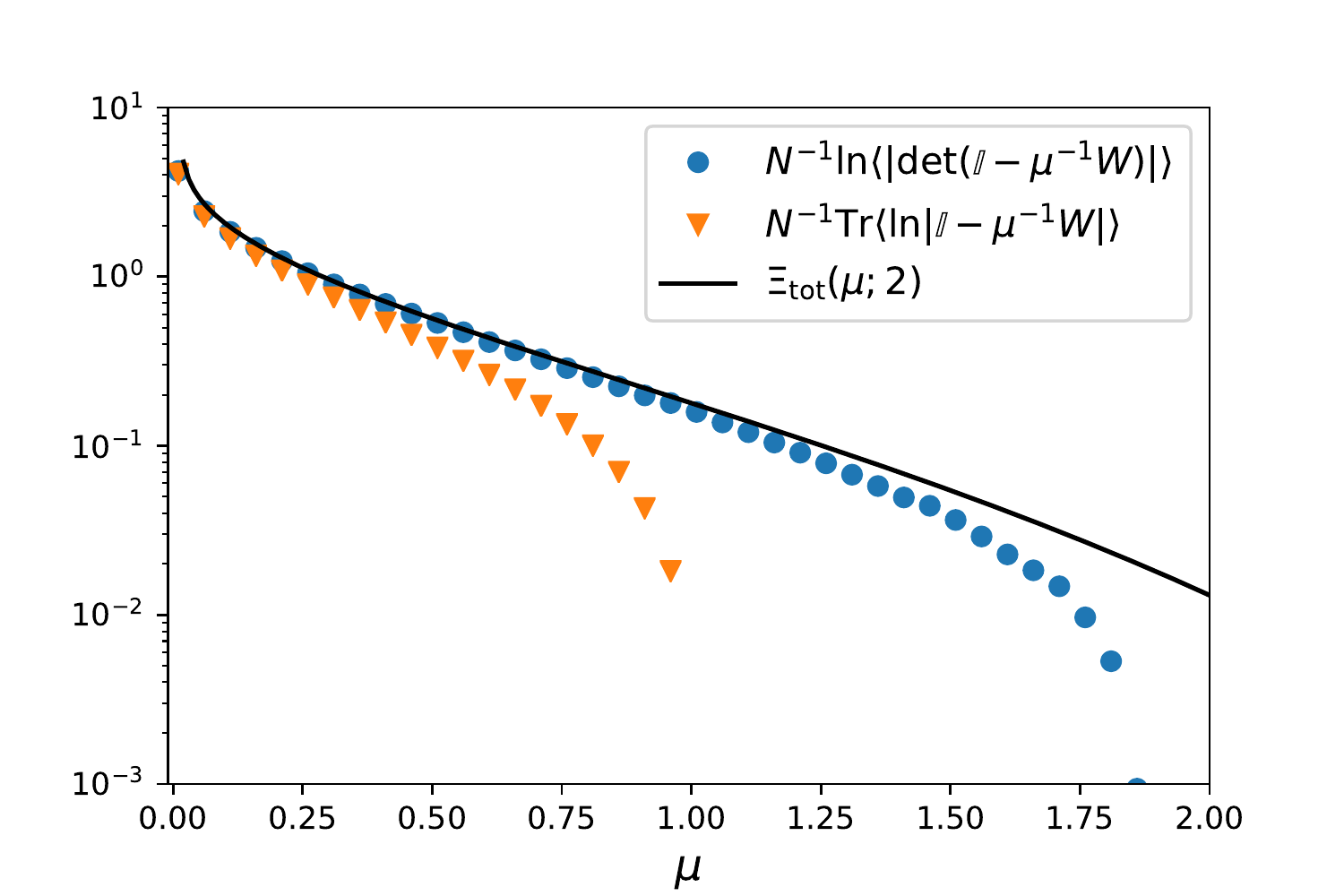} 
\caption{Comparison between the annealed total complexity $\Xi_{\rm tot}(\mu;\alpha)$ for $\alpha=2$ (black solid line) obtained by inserting the solution of Eqs.(\ref{sp_sol_1_res}-\ref{sp_sol_2_res}) into Eq.\eqref{Xi_tot_res} and numerical simulations of $N^{-1}\ln\moy{\frac{|\det(\mu\mathbb{I}-W)|}{\mu^N}}$ (blue circle) and $N^{-1}\moy{\Tr\ln\frac{|\det(\mu\mathbb{I}-W)|}{\mu^N}}$ (orange triangles) for $N=200$, plotted as a function of $\mu$. It is expected that these three quantities coincide in the $N\to \infty$ limit. The agreement is good for low values of $N$ but there is some discrepancy for larger values of $\mu$. This discrepancy results both from finite size effects and from the fact that the average is dominated by rare events, preventing an efficient numerical evaluation.}
\label{Fig_Xi_tot}
\end{figure}

Our main result is that in the limit $M,N\to \infty$ with fixed ratio $\alpha=M/N>1$ the total annealed complexity for the landscape \eqref{landscape} with the random potential \eqref{mod_superpos} can be expressed  as 
\begin{align}\label{Xi_tot_res}
\Xi_{\rm tot}(\mu;\alpha)&=\lim_{N\to \infty}\frac{1}{N}\ln \moy{{\cal N}_{\rm
 tot}(\mu;\alpha)}=\int_{\mu}^{\infty}\left(\frac{1}{\nu}-m_r(\nu)\right)\,d\nu\;.  
\end{align}
in terms of a function $m_r(\mu):=m_r$ which together with its counterpart $m_i(\mu):=m_i$ satisfies the pair of equations:
\begin{align}
   \mu\int_{-\infty}^{\infty} dt\,e^{-\frac{t^2}{2}}\sqrt{t^2 m_i^2+(1-t\,m_r)^2}\label{sp_sol_1_res}&=\alpha\,\int_{-\infty}^{\infty} dt\,\frac{t\,e^{-\frac{t^2}{2}}}{\sqrt{t^2 m_i^2+(1-t\,m_r)^2}}\,,
   \\
   \int_{-\infty}^{\infty} dt\,e^{-\frac{t^2}{2}}\sqrt{t^2 m_i^2+(1-t\,m_r)^2}\label{sp_sol_2_res}&=\alpha\,(m_i^2+m_r^2)\int_{-\infty}^{\infty} dt\,\frac{t^2\,e^{-\frac{t^2}{2}}}{\sqrt{t^2 m_i^2+(1-t\,m_r)^2}}\,.
\end{align}

Using this expression we show that in sharp contrast to previously studied models with isotropic harmonic confinement, the total annealed complexity does not undergo a "topology trivialisation transition" and stays positive for any finite value $0<\mu<\infty$ of the strength of the confinement. Still, it decreases very rapidly as $\mu\to \infty$:
\be
\Xi_{\rm tot}(\mu;\alpha)\approx \frac{\alpha}{\mu^3} \sqrt{\frac{2}{\pi}}e^{-\frac{\mu^2}{2}}\;,\;\;\mu\to \infty\;.\label{xi_tot_mu_large_res}
\ee

It is worth noting that methodologically the main difference with previously studied cases of Eq.\eqref{trans_rot_inv} is that the annealed complexity is obtained using the functional variational principle involving the mean spectral density $n(t)$ of the eigenvalues of the random matrix $T$ instead of a variation principle over a single real random variable as in \cite{F04} and related studies.

\subsection{Results on the general Marchenko-Pastur Ensemble}

In this section we give the main outcomes of our studies of spectral characteristics of general matrices of the form $W=K T K^{T}$ related to the Marchenko-Pastur ensemble,
not only those directly involved in the computation of annealed complexity.\\[1ex] 

\noindent {\it Mean asymptotic eigenvalue density for GMPE}: \\[0.5ex]
Consider the mean spectral density $\rho(\lambda)=\lim_{N\to \infty}\frac{1}{N}\sum_{k=1}^{N}
\moy{\delta(\lambda-\lambda_k)}$ for the Gaussian MPE matrices $W$ defined in \eqref{GMPE_Matrix}.
It turns out to be convenient to define the associated resolvent/Stiltjes transform, defined here for a real spectral parameter $\lambda$ as
\begin{align}
m(\lambda)&=\int_{-\infty}^{\infty} d\lambda'\frac{\rho(\lambda')}{\lambda-\lambda'}=\dashint_{-\infty}^{\infty} d\lambda'\frac{\rho(\lambda')}{\lambda-\lambda'}+i\pi \rho(\lambda):=m_r(\lambda)+i\, m_i(\lambda)\;,   
\end{align}
where $m_r\equiv m_r(\lambda)$ and $m_i\equiv m_i(\lambda)$ are respectively its real and imaginary parts. The mean spectral density can be expressed in terms of the resolvent $m(\lambda)$ as $\rho(\lambda)=\frac{\Im[m(\lambda)]}{\pi}$ where for GMPE the resolvent  $m(\lambda)$ is obtained by solving the following transcendental equation:
\begin{equation}\label{MM_res}
\lambda=\frac{1-\alpha}{m(\lambda)}+\frac{i\alpha}{m^2(\lambda)}\Psi\left(\frac{1}{i\, m(\lambda)}\right)\;,
\end{equation}
with $\Psi(u)=\sqrt{\frac{\pi}{2}}e^{\frac{u^2}{2}}(1+\erf(u/\sqrt{2}))$. The resulting mean spectral density can be shown to be even: $\rho(\lambda)=\rho(-\lambda)$ and supported on the whole real line, with asymptotic behaviour as $\lambda\to \pm\infty$ given by
\be
\rho(\lambda)\approx \frac{\alpha}{\sqrt{2\pi}}e^{-\frac{\lambda^2}{2}}\;,\;\;\lambda\to \pm \infty\;.
\ee
For $\alpha> 1$, the density $\rho(\lambda)$ takes its finite maximum value at the origin $\lambda=0$, where $\rho_\alpha:=\rho(\lambda=0)$ is the solution of the transcendental equation obtained by inserting $m(0)=i\pi \rho_{\alpha}$ in Eq. \eqref{MM_res}. For $\alpha=1$ the density diverges close to the origin as $\rho(\lambda)\approx 1/\sqrt{(2\pi)^{3/2}|\lambda|}$.

\begin{figure}[t!]
\centering
 \includegraphics[width=.8\textwidth]{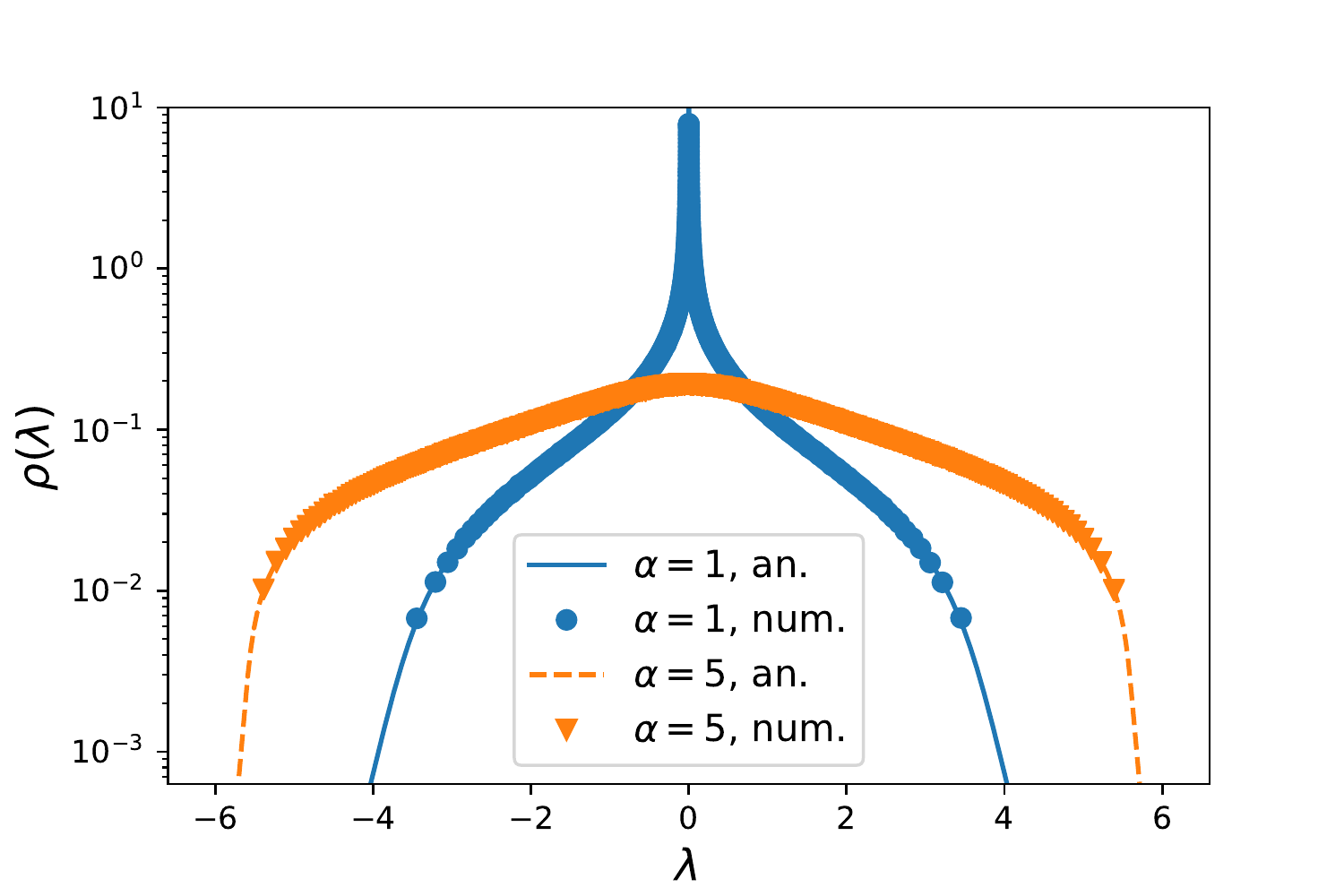} 
\caption{Comparison between the average density $\rho(\lambda)$ for $\alpha=1,5$ obtained analytically in the limit $N\to\infty$ from Eq. \eqref{MM_res} by using that $\rho(\lambda)=\frac{\Im[m(\lambda)]}{\pi}$ (respectively blue solid and orange dashed lines) and numerically for a matrix of finite size $N=500$ (respectively blue circle and orange triangle).}
\label{Fig_rho}
\end{figure}

Let us finally mention that for $\alpha<1$ the density $\rho(\lambda)$ displays a delta peak at the origin $(1-\alpha)\delta(\lambda)$ reflecting the presence of $M-N$ zero eigenvalues.
At the same time its continuous part tends to a finite value at the origin: $\alpha/(\sqrt{2\pi}(1-\alpha))$.
In Fig. \ref{Fig_rho}, we plot a comparison between our analytical prediction for the average density of eigenvalue $\rho(\lambda)$ in the $N\to \infty$ limit and results from numerical simulation for $N=500$, showing excellent agreement.

\vspace{0.2cm}

\noindent {\it Moments and correlation functions of characteristic polynomials}: \\[0.5ex]
For a fixed matrix $T$ the positive moments of the characteristic polynomial ${\cal Z}_{\beta,p}(\lambda;T)\equiv {\cal Z}_{\beta,p}(\lambda{\bm 1};T)$ defined in Eq.\eqref{charpol} with setting ${\bm \lambda}=\lambda{\bm 1}=\lambda(1,\cdots,1)$ can be expressed for any integer $p>0$ and any size $N$  in terms of a $p\times p$ determinant for $\beta=2$
\be
{\cal  Z}_{\beta=2,p}(\lambda;T)=\frac{\det\Big[g_{N,k-j}(\lambda;{ T})\Big]_{j,k=1}^{p} }{\det\Big[q_{N,k-j}\Big]_{j,k=1}^p}
\ee
and in terms of a $2p\times 2p$ Pfaffian for $\beta=1$
\begin{equation}
\mathcal{Z}_{\beta=1,p}(\mathbf{\lambda};T)=
\frac{\Pf\Big[(j-k)g_{N,2p+1-(k+j)}(\lambda;{ T})\Big]_{j,k=1}^{2p}}{\Pf\Big[(j-k)q_{N,2p+1-(k+j)}\Big]_{j,k=1}^{2p}}\;.
\end{equation}
In these expressions, the functions $g_{N,m}(\lambda;{ T})$ and $q_{N,m}$ are defined as
\begin{align}
    &g_{N,m}(\lambda;{T})=\lambda^{N+m}\sum\limits_{l=0}^{N+m}\frac{(-1)^l}{(N+m-l)!}e_{l}\left(\frac{T}{N\lambda}\right)\;,\;\;q_{N,m}=g_{N,m}(1;\mathbb{O})=\frac{1}{(N+m)!}\;,
\end{align}
where  $e_l(X_1,\cdots,X_n)=\sum_{1\leq i_1<i_2<\cdots<i_l\leq n}\prod_{j=1}^l X_{i_j}$ is the $l^{\rm th}$ elementary symmetric polynomial. Note that taking $T=\mathbb{I}$, one recovers the known results for the Wishart-Laguerre ensemble, see e.g. \cite{Shch2}.

For a random diagonal matrix $T$ (not necessarily normally-distributed) we only managed to consider the mean of the characteristic polynomials and its two-point correlation function, $p=2$. We find that the mean characteristic polynomial only depends on the first moment $\moy{T}=\int dt\,t\,n_M(t)$, where $n_M(t):=\frac{1}{M}\sum_{a=1}^M\delta(t-T_a)$ is the associated density function. For non-zero $\moy{T}\neq 0$,
\begin{align}
\mathcal{Z}_{\beta,1}(\lambda)&=\moy{\mathcal{Z}_{\beta,1}(\lambda;T)}_T=\bigg(\frac{\moy{T}}{N}\bigg)^N\frac{M!}{(M-N)!} {}_1F_1\bigg(-N;1+M-N;\frac{\lambda N}{\moy{T}}\bigg)\;,
\end{align}
in terms of the hypergeometric function  defined as ${}_pF_q({\bm a};{\bm b};x)=\sum_{n=0}^{\infty}\frac{\prod_{i=1}^p (a_i)_n}{\prod_{j=1}^q (b_j)_n\, n!}x^n$. For $\moy{T}=0$ we instead find $\mathcal{Z}_{\beta,1}(\lambda)=\lambda^N$. In the large $N$ limit, we study the rate of growth of the mean characteristic polynomial and show that
\begin{align}
\Xi_{1}(x;\alpha)&=\lim_{N\to \infty}\frac{1}{N}\ln\left|\frac{\mathcal{Z}_{\beta,1}(\lambda)}{\lambda^N}\right|=\begin{cases}
{\cal L}_1(w_+;x,\alpha)&\;,\;\;x<x_-(\alpha)\;, \\
&\\
\Re\left[{\cal L}_1(w_\pm;x,\alpha)\right]&\;,\;\;x_-(\alpha)\leq x\leq x_+(\alpha)\\
&\\
{\cal L}_1(w_-;x,\alpha)&\;,\;\;x>x_+(\alpha)\;,
\end{cases}
\end{align}
where we denoted $x=\lambda/\moy{T}$, $x_{\pm}(\alpha)=(1\pm\sqrt{\alpha})^2$ and introduced 
\begin{align}
{\cal L}_1(w;x,\alpha)&=x w-\ln x w+\alpha\ln\left(1-w\right)-1\;,\\
w_{\pm}(x,\alpha)&=\frac{1-\alpha+x\pm\sqrt{(\alpha-1-x)^2-4x}}{2x}\;.
\end{align}

The two-point spectral correlation function of the characteristic polynomials has been computed for any size $N$ in the case of vanishing mean $\moy{T}=0$. It then turns out to depend on $T$ only through its variance/second moment $\moy{T^2}=\int dt\,t^2\,n(t)$ and reads
\begin{align}\label{Z_b_2}
&\mathcal{Z}_{\beta,2}(\bm{\lambda}=(\lambda_1,\lambda_2))=\\
&\bigg(\frac{\moy{T^2}}{N^2}\bigg)^N\frac{\beta}{2}\Gamma\left(N+\frac{2}{\beta}+1\right)\frac{M!}{(M-N)!}{}_1F_2\bigg(-N;\frac{2}{\beta}+1,1+M-N;-\frac{ \lambda_1\lambda_2 N^2}{\moy{T^2}}\bigg)\;.\nn
\end{align}
 In particular in the large $N$ limit we define
 \be
\Xi_{2}(u;\alpha)=\lim_{N\to \infty}\frac{1}{N}\ln\left|\frac{\mathcal{Z}_{\beta,2}({\bm \lambda})}{(\lambda_1\lambda_2)^N}\right|={\cal L}_2(\eta_*;u,\alpha)\;,
\ee
and find that it only depends on the rescaled positive parameter $u=|\lambda_1\lambda_2|/\moy{T^2}\geq 0$:
 \begin{align}
{\cal L}_2(\eta;u,\alpha)=&\alpha\ln(\alpha)-(\alpha+\eta-1)\ln (\alpha+\eta-1)\\
&+\eta\ln (u(1-\eta ))+2\eta-2\eta\ln(\eta)-\ln (1-\eta)-2-\ln u\nn\;,\\
\eta_*=\frac{1-\alpha }{3}&-\frac{\sqrt[3]{2} \Theta}{3 \sqrt[3]{\Delta +\sqrt{\Delta ^2+4 \Theta^3}}}+\frac{\sqrt[3]{\Delta+\sqrt{\Delta ^2+4 \Theta^3}}}{3 \sqrt[3]{2}}\;,\nn\\
{\rm where}\;\;\Delta=&9u(2+\alpha)-2(\alpha-1)^3\;,\;\;\Theta=3 u-(\alpha -1)^2\;.
\end{align}

Note a somewhat surprising feature of the two-point function in the Marchenko-Pastur ensembles with random matrix $T$: it depends only on the product $\lambda_1\lambda_2$, in sharp contrast with earlier studied cases  \cite{BH1} where in the large-$N$ limit it rather depended on the spectral difference $|\lambda_1-\lambda_2|$.

\begin{acknowledgments}
 Submitting this article to the collection celebrating  Freeman Dyson, one of the founding fathers of the modern random matrix theory, we would also like to dedicate it to Vladimir Marchenko who turns 100 years old in July 2022, and whose pioneering 1967 paper with Leonid Pastur deeply influenced subsequent development of the field.

We would like to thank Pierre Le Doussal for his interest in the work, and in particular for pointing out a few relevant references.
 The research by Bertrand Lacroix A Chez Toine and Yan V Fyodorov was supported by the EPSRC Grant EP/V002473/1 {\it Random Hessians and Jacobians: theory and applications}.
Sirio Belga Fedeli acknowledges support by the EPSRC Centre for Doctoral Training
in Cross-Disciplinary Approaches to Non-Equilibrium Systems through Grant No. EP/L015854/1.
\end{acknowledgments}

\section{Solution of the annealed counting problem}\label{sec_count}

In this section we show how to solve the annealed counting problem discussed in the introduction. The starting point of our consideration is the standard Kac-Rice formula for the total number of equilibria which in the case \eqref{landscape} reads
\begin{align}
&\moy{{\cal N}_{\rm tot}(\mu;\alpha)}=\label{KR_N_tot}\int_{\mathbb{R}^N} d^N{\bf x} \,\moy{\prod_{i=1}^N \delta(\mu\,x_i+\partial_{x_i}V({\bf x}))|\det(\mu \mathbb{I}-W({\bf x}))|}\;,
\end{align}
where $W_{ij}({\bf x})=-\partial_{x_i,x_j}V({\bf x})$ and the expectation $\moy{\cdots}$ in the case of potential \eqref{mod_superpos} has been taken over both the components of the vector ${\bf k}_{a}$, and the coefficients $u_{1,2}^{(a)}$ for $a=1,\cdots,M$. It turns out to be expedient to pass from the original random variables $u_{1,2}^{(a)}, \, a=1,\cdots,M$, to the new set of  random variables obtained by rotation as
\begin{align}
T_a&=u_1^{(a)}\cos({\bf k}_a^{T}\cdot{\bf x})+u_2^{(a)}\sin({\bf k}_a^{T}\cdot{\bf x})\;,\\
G_a&=u_1^{(a)}\sin({\bf k}_a^{T}\cdot{\bf x})-u_2^{(a)}\cos({\bf k}_a^{T}\cdot{\bf x})\;.
\end{align}
Despite the rotation used to define $T_a$ and $G_a$ being explicitly ${\bf x}$ dependent, the resulting random variables are statistically ${\bf x}$ independent, with zero mean and covariance
\be
\moy{T_a T_b}=\moy{G_a G_b}=\delta_{a,b}\,,\,\,\moy{T_a G_b}=0\,,\,\,a,b=1,\cdots,M\,.
\ee
  The random force $f_i=-\partial_{x_i}V({\bf x})$ and the Hessian $W_{ij}=-\partial_{x_i,x_j}V({\bf x})$ associated to the potential $V({\bf x})$ are conveniently expressed in terms of these random variables as
\begin{align}
f_i&\equiv -\partial_{x_i}V({\bf x})=\sum_{a=1}^M k_{a,i} G_a\;,\\
W_{ij}&\equiv-\partial_{x_i,x_j}V({\bf x})=\sum_{a=1}^M k_{a,i}k_{a,j} T_a\;,
\end{align}
and using the properties of $G_a,T_a$ and $K$, it is easy to see that these random variables are also independent of the position ${\bf x}$. Now we may use in Eq. \eqref{KR_N_tot} the Fourier integral representation for the Dirac delta function
\begin{align}
\delta(\mu\, x_j-f_j)&=\int_{-\infty}^{\infty} \frac{dq_j}{2\pi}e^{i\,q_j(\mu\,x_j-f_j)}=\int_{-\infty}^{\infty} \frac{dq_j}{2\pi}e^{i\,q_j(\mu\,x_j-\sum_{a=1}^M k_{a,j}\,G_a)}, 
\end{align}
which allows to take explicitly the expectation with respect to the random variables $G_a$'s yielding
\begin{align}
    \moy{{\cal N}_{\rm tot}(\mu;\alpha)}=&\int_{\mathbb{R}^N} d^N{\bf x} \,\int_{\mathbb{R}^N} \frac{d^N {\bf q}}{(2\pi)^N}e^{i\mu \sum_{j=1}^N q_j\,x_j}\moy{ e^{-\sum_{a=1}^M \frac{(k_{a}^T\cdot {\bf q})^2}{2}}|\det(\mu\,\mathbb{I}-K T K^T)|}_{K,T}\;.
\end{align}
Further integration over ${\bf x}$ yields
\begin{align}
\moy{{\cal N}_{\rm tot}(\mu;\alpha)}=&\int_{\mathbb{R}^N} d^N {\bf q}\prod_{j=1}^{N}\delta(\mu\,q_j)\moy{ e^{-\sum_{a=1}^M \frac{(k_{a}^T\cdot {\bf q})^2}{2}}|\det(\mu\,\mathbb{I}-K T K^T)|}_{K,T}\;,\\
=&\frac{\moy{|\det(\mu\,\mathbb{I}-K T K^T)|}_{K,T}}{\mu^N}\;,\nn
\end{align}
and after writing explicitly the expectation over the diagonal matrix $T$ the total number of equilibria finally reads
\begin{align}
&\moy{{\cal N}_{\rm tot}(\mu;\alpha)}=\label{ntot_sirio}\frac{1}{\mu^N}\int_{\mathbb{R}^M} \prod_{a=1}^M \frac{dT_a}{\sqrt{2\pi}}\, e^{-\sum_{a=1}^M\frac{T_a^2}{2}}\moy{|\det(\mu\mathbb{I}-K T K^{T})|}_K\;,
\end{align}
where $\moy{\cdots}_K$ is the remaining expectation over the Gaussian random matrix $K$. 

We will now focus on the limit $N,M \to \infty$ with $1\leq M/N=\alpha<\infty$. In this particular limit, we define the annealed total complexity as
\begin{align}\label{xi_tot_abs_det}
\Xi_{\rm tot}(\mu;\alpha)&=\lim_{N\to \infty}\frac{1}{N}\ln \moy{{\cal N}_{\rm tot}(\mu;\alpha)}=\lim_{N\to \infty}\frac{1}{N}\ln\frac{\moy{|\det(\mu\, \mathbb{I}-W)|}}{\mu^N}\;,
\end{align}
where we remind that $W=K T K^T$. In order to make progress in this computation, we assume that for a fixed diagonal matrix $T$ holds the strong self-averaging property, i.e.
\begin{align}
&\lim_{N\to \infty}\frac{1}{N}\ln\moy{|\det(\mu \mathbb{I}-KTK^{T})|}_{K}\label{SA_hyp}=\lim_{N\to \infty}\frac{1}{N}\Tr\moy{\ln|\mu \mathbb{I}-KTK^{T}|}_{K}=\int d\lambda\, \rho(\lambda)\,\ln|\lambda-\mu| \;,
\end{align}
where the mean limiting spectral density $\rho(\lambda)$ is defined via
\be
\rho(\lambda)=\lim_{N\to \infty} \frac{1}{N}\sum_{i=1}^{N}\moy{\delta(\lambda-\lambda_i)}_K\;,
\ee
with the $\lambda_i$'s being the random eigenvalues of $W=K T K^T$. Note that in the model with Gaussian random potential satisfying Eq.\eqref{trans_rot_inv} a similar strong self-averaging property has been proven rigorously \cite{ABAC13a,ABAC13b} in considerable generality. With the assumption of its validity for the present model we may express the average total number of equilibria/stationary points for finite but large $N$ as
\begin{align}
&\moy{{\cal N}_{\rm tot}(\mu;\alpha)}\approx\frac{1}{\mu^N}\int_{\mathbb{R}^M} \prod_{a=1}^M \frac{dT_a}{\sqrt{2\pi}}e^{\displaystyle N\left(-\frac{\alpha}{M}\sum_{a=1}^M\frac{T_a^2}{2}+\int_{-\infty}^{\infty} d\lambda\, \rho(\lambda)\,\ln|\lambda-\mu|\right)}\;. 
\end{align}
To facilitate extracting the large-$N$ asymptics we find it expedient to introduce the normalised limiting density of $T_a$ as
\be
n(t)=\lim_{M\to \infty} \frac{1}{M}\sum_{a=1}^{M}\delta(T_a-t)\
\ee
and replace the $M$-dimensional integral over the components $T_a$'s by a functional integral over this density
as
\begin{align}
\moy{{\cal N}_{\rm tot}(\mu;\alpha)}\approx &\frac{\displaystyle \int_{-\infty}^{\infty} dl \int {\cal D}n(t)e^{-\displaystyle N\,S_{\rm tot}\left[n;\mu,\alpha,l\right]}}{\displaystyle \mu^N\int_{-\infty}^{\infty} dl \int {\cal D}n(t)e^{-\displaystyle \alpha\,N\,S_{\rm den}\left[n;l\right]}}\;,\\
S_{\rm den}\left[n;l\right]=&\frac{1}{2}\int_{-\infty}^{\infty} dt\,n(t)\,t^2+\int_{-\infty}^{\infty}\,dt\,n(t)\,\ln n(t)+l\left(\int_{-\infty}^{\infty} dt\,n(t)-1\right)\;,\\
S_{\rm tot}\left[n;\mu,\alpha,l\right]=&\alpha\,S_{\rm den}\left[n;l\right]-\int_{-\infty}^{\infty} d\lambda\,\rho(\lambda)\ln|\mu-\lambda|\;.
\end{align}
where $l$ is the Lagrange multiplier necessary to ensure the correct normalization of the density. 
This form is now amenable to using the Laplace method justified by the large parameter $N$, yielding for the annealed complexity
\begin{align}\label{complexity_opti}
\Xi_{\rm tot}(\mu;\alpha)=&\lim_{N\to \infty}\frac{1}{N}\ln \moy{{\cal N}_{\rm tot}(\mu;\alpha)}\alpha S_{\rm den}\left[n_{\rm den};l_{\rm den}\right]-S_{\rm tot}\left[n_{\rm tot};\mu,\alpha,l_{\rm tot}\right]-\ln \mu\;,
\end{align}
where we have denoted $n_{\rm den},n_{\rm tot}$ and $l_{\rm tot},l_{\rm den}$ the values of the limiting density and Lagrange parameters that minimise $S_{\rm den},S_{\rm tot}$ respectively. Note that the whole procedure is not unlike to what has been done when evaluating the complexity for Gaussian translationally and rotationally invariant model in \cite{F04}, however instead of maximising with respect to a single real variable one has to optimize with respect to the limiting density $n(t)$ in the present case. We also note that although the optimization is presented here in an informal style of theoretical physics, we believe it can be recast rigorously using the Large Deviations framework.

Let us first consider the stationarity conditions for the action functional $S_{\rm den}$. Requiring its variation with respect to $n(t)$ and the derivative with respect to the Lagrange parameter to be zero, one obtains the equations
\begin{align}
\left.\frac{\delta S_{\rm den}}{\delta n(t)}\right|_{n=n_{\rm den},l=l_{\rm den}}&=\frac{t^2}{2}+l_{\rm den}+1+\ln n_{\rm den}(t)=0\;,\\
\left.\frac{\partial S_{\rm den}}{\partial l}\right|_{n=n_{\rm den},l=l_{\rm den}}&=\int dt\,n_{\rm den}(t)-1=0\;.
\end{align}
One can check easily that the distribution $n_{\rm den}(t)$ is then given by the properly normalized Gaussian density
\be
n_{\rm den}(t)=\frac{e^{-\frac{t^2}{2}}}{\sqrt{2\pi}}\;,
\ee
 for which $S_{\rm den}\left[n_{\rm den};l_{\rm den}\right]=-l_{\rm den}-1=-\frac{1}{2}\ln(2\pi)$.

Similarly, varying the action functional $S_{\rm tot}$ with respect to $n(t)$ yields the stationarity condition
\begin{align}
\left.\frac{\delta S_{\rm tot}}{\delta n(t)}\right|_{n=n_{\rm tot}}=&\alpha\left[\frac{t^2}{2}+l_{\rm tot}+1+\ln n_{\rm tot}(t)\right]\label{sp_eq}-\int_{-\infty}^{\infty} d\lambda\, \left.\frac{\delta \rho(\lambda)}{\delta n(t)}\right|_{n=n_{\rm tot}}\ln|\mu-\lambda|=0\;, 
\end{align}
while the differentiation with respect to the Lagrange parameter $l$ ensures again the normalisation for $n_{\rm tot}(t)$. To find $n_{\rm tot}(t)$ and the associated action value one needs to compute explicitly the functional derivative of $\rho(\lambda)$ with respect to $n(t)$. We will now show that for the specific type of matrix considered here, this problem is exactly solvable. The resolvent/Stiltjes transform function $m(z)$ associated with the density $\rho(\lambda)$ satisfies for a fixed limiting density $n(t)$ a self-consistency equation derived by Marchenko and Pastur in \cite{MP67} which reads
\be
m(z)=\int_{-\infty}^{\infty} d\lambda\,\frac{\rho(\lambda)}{z-\lambda}=\left(z-\alpha\int_{-\infty}^{\infty} dt\,\frac{t\,n(t)}{1-t\,m(z)}\right)^{-1}\;.\label{res_eq}
\ee
The resolvent can be introduced in the saddle-point equation \eqref{sp_eq} using the identity, 
\begin{align}
\int_{-\infty}^{\infty} d\lambda\,\rho(\lambda)\,(\ln|\mu-\lambda|-\ln|\mu|)&=\int_{\infty}^{\mu}d\nu\,\left[\Re\left(\int_{-\infty}^{\infty} d\lambda\,\frac{\rho(\lambda)}{\nu-\lambda}\right)-\frac{1}{\nu}\right]\\
&=\int_{\mu}^{\infty}d\nu\,\left(\frac{1}{\nu}-m_r(\nu)\right)\;, \nn
\end{align}
where $m_r(\nu)$ is the real part of the resolvent $m(\nu)$ at the real spectral parameter $\nu$ and we have used that $z\, m_r(z)=z\dashint_{-\infty}^{\infty}d\lambda\,\rho(\lambda)/(z-\lambda)\to 1$ for any real or complex spectral parameter $z$ in the limit $|z|\to \infty$. Using the self-consistency equation \eqref{res_eq}, one can compute  the functional derivative of $m(z)$ at any complex spectral parameter $z$ with respect to $n(t)$ explicitly:
\begin{align}
\frac{\delta m}{\delta n(t)}=&\left[1-\alpha\, m^2\int_{-\infty}^{\infty} dt'\,\frac{{t'}^2\,n(t')}{(1-t'\,m)^2}\right]^{-1}\frac{\alpha\,t\,m^2}{1-t\, m}\;.
\end{align}
Comparing this expression with the derivative of the resolvent $m(z)$ with respect to $z$,
\begin{align}
\partial_z m=&-\left[1-\alpha\, m^2\int_{-\infty}^{\infty} dt'\,\frac{{t'}^2\,n(t')}{(1-t'\,m)^2}\right]^{-1} m^2\;,
\end{align}
one obtains a simple identity
\be
\frac{\delta m}{\delta n(t)}=-\frac{\alpha\,t}{1-t\, m}\partial_z m=\alpha\,\partial_z\ln[1-t\,m]\;.
\ee
This identity can then be used to obtain
\begin{align}
\frac{\delta}{\delta n(t)}\int_{-\infty}^{\infty} d\lambda\,\rho(\lambda)\,(\ln|\mu-\lambda|-\ln\mu)&=\int_{\infty}^{\mu}d\nu \frac{\delta}{\delta n(t)}\left[\Re\left(\int_{-\infty}^{\infty} d\lambda\,\frac{\rho(\lambda)}{\nu-\lambda}\right)-\frac{1}{\nu}\right]\\
&=\alpha\,\Re\left(\int_{\infty}^{\mu}d\nu\,\partial_{\nu}\ln\left[1-t\,m(z)\right]\right)\nn\\
&=\frac{\alpha}{2}\ln[(1-t\,m_r(\mu))^2+t^2 m_i(\mu)^2]\;,  \nn  
\end{align}
where we introduced $m_r(\mu)$ and $m_i(\mu)$ as the real and imaginary parts of the resolvent $m(\mu)$ at the real spectral parameter $^\mu$, respectively. Now the stationarity condition \eqref{sp_eq} implies that the optimal density $n_{\rm tot}(t)$ satisfies the equation
\be
\frac{t^2}{2}+l_{\rm tot}+1+\ln n_{\rm tot}=\frac{1}{2}\ln[(1-t\,m_r(\mu))^2+t^2 m_i(\mu)^2]\;
\ee
which is immediately solved in terms of the real and imaginary parts of the resolvent at position $\mu$ yielding
\be
n_{\rm tot}(t)=\frac{e^{-\frac{t^2}{2}}}{\sqrt{2\pi} Z(\mu)}\sqrt{(1-t\,m_r(\mu))^2+t^2 m_i(\mu)^2}\;,
\ee
with $Z(\mu)=e^{l_{\rm tot}+1}/\sqrt{2\pi}=e^{l_{\rm tot}-l_{\rm den}}$ being the $\mu-$dependent normalisation factor given explicitly by 
\be
Z(\mu)=\int_{-\infty}^{\infty}dt\,\frac{e^{-\frac{t^2}{2}}}{\sqrt{2\pi}}\sqrt{(1-t\,m_r(\mu))^2+t^2 m_i(\mu)^2}\;.\label{Z_mu_1}
\ee
Finally, at a real value of $\mu$ the self-consistency equation \eqref{res_eq} for the full resolvent $m(\mu)$ can be rewritten as a pair of equations for the real and imaginary parts $m_r\equiv m_r(\mu)$ and $m_i\equiv m_i(\mu)$:
\begin{align}
\frac{m_r}{m_i^2+m_r^2}&=\mu-\alpha\int_{-\infty}^{\infty} dt\,\frac{t\,n_{\rm tot}(t)\,(1-t\,m_r)}{m_i^2\,t^2+(1-t\,m_r)^2}\,,\label{m_r_eq_1}\\
\frac{m_i}{m_i^2+m_r^2}&=\alpha\int_{-\infty}^{\infty} dt\,\frac{t^2\,n_{\rm tot}(t)\,m_i}{m_i^2\,t^2+(1-t\,m_r)^2}\,.\label{m_i_eq_1}
\end{align}
Recall that the imaginary part of the resolvent is expressed in term of the limiting spectral density as $m_i(\mu)=\pi \rho(\mu)$. For the optimal density $n_{\rm tot}(t)$ of $T$ being unbounded, it is expected that the limiting density $\rho(\lambda)$ of $W=K T K^{T}$ is also unbounded, see \cite{MP67}. Hence $m_i(\mu)>0$ for any $0<\mu<\infty$ and one can divide by $m_i$ on both sides of Eq. \eqref{m_i_eq_1}. Inserting then the expression of $n_{\rm tot}(t)$ one obtains the more explicit equations (\ref{sp_sol_1_res}-\ref{sp_sol_2_res}). This pair of equations allows to express $m_i(\mu)$ and $m_r(\mu)$ at the optimal solution. One now needs to relate these values to the complexity expression \eqref{complexity_opti} for $\Xi_{\rm tot}(\mu;\alpha)$. In order to obtain a convenient representation for the total complexity we first compute its derivative with respect to $\mu$ as
\begin{align}
\partial_\mu \Xi_{\rm tot}&=\dashint_{-\infty}^{\infty} d\lambda\,\frac{\rho(\lambda)}{\mu-\lambda}-\frac{1}{\mu}-\partial_\mu n_{\rm tot}\left. \frac{\delta S_{\rm tot}}{\delta n(t)}\right|_{n=n_{\rm tot}}=m_r(\mu)-\frac{1}{\mu}\;,
\end{align}
where we have used that $n_{\rm tot}$ is the density that minimises $S_{\rm tot}$.
Finally, using again that as $\mu\to \infty$, $m_r(\mu)=\dashint d\lambda\,\rho(\lambda)/(\mu-\lambda)\to \mu^{-1}$ such that $m_r(\mu)-\mu^{-1}\to 0$, one arrives at the final expression for the annealed complexity:
\be
\Xi_{\rm tot}(\mu;\alpha)=\int_{\mu}^{\infty}d\nu\left(\frac{1}{\nu}-m_r(\nu)\right)\;.
\ee
In the limit $\mu\to \infty$, we expect that $m_r(\mu)\sim \mu^{-1}$ while $m_i(\mu)\propto e^{-\mu^2/2}$ which is inherited from the Gaussian decay of $n_{\rm tot}(t)$ (see also a related discussion on the asymptotic of spectral density below). In that limit, one can safely set $m_i(\mu)\to 0$ in Eqs. \eqref{Z_mu_1} and \eqref{m_r_eq_1} which yields 
\begin{align}
Z(\mu)&\approx \int_{-\infty}^{\infty}dt\,\frac{e^{-\frac{t^2}{2}}}{\sqrt{2\pi}}|1-t\,m_r(\mu)|=\sqrt{\frac{2}{\pi}}m_r\,e^{-\frac{1}{2m_r^2}}+\erf\left(\frac{1}{\sqrt{2}m_r}\right)\\
&\approx 1+\sqrt{\frac{2}{\pi}}m_r^3 e^{-1/m_r^2} \;,\;\;\mu\to \infty\;,\nn\\
\frac{1}{m_r(\mu)}&\approx \mu-\alpha\int_{-\infty}^{\infty} dt\, \frac{t\,e^{-\frac{t^2}{2}}}{\sqrt{2\pi}Z(\mu)}{\rm sign}(1-t\,m_r(\mu))\mu+\alpha \sqrt{\frac{2}{\pi}}e^{-1/(2m_r^2)}\;,\;\;\mu\to \infty\;.
\end{align}
In particular, from the second equation, one obtains that 
\begin{align}
\Xi_{\rm tot}(\mu;\alpha)&=\int_{\mu}^{\infty}d\nu \left(\frac{1}{\nu}-m_r(\nu)\right)\approx \alpha \sqrt{\frac{2}{\pi}} \int_{\mu}^{\infty}\frac{d\nu}{\nu^2} e^{-\nu^2/2}\approx \frac{\alpha}{\mu^3} \sqrt{\frac{2}{\pi}}e^{-\frac{\mu^2}{2}}\;,\;\;\mu\to \infty\;,\label{xi_tot_mu_large} 
\end{align}
showing that the total complexity is never zero for any finite strength of the isotropic confinement $\mu$, although it vanishes very rapidly at large $\mu$.
Such behaviour is in stark contrast with previously considered models with isotropic confinement, where the complexity was found to vanish for $\mu>\mu_c$, signalling of the total  "landscape topology trivialization'' beyond a finite value of the confining parameter.

\section{Spectral density of the Gaussian Marchenko-Pastur Ensemble} \label{sec_dens}
In this section we study the simplest spectral characteristic of GMPE, its mean spectral density in the limit $N,M\to \infty$ with $\alpha=M/N=O(1)$ defined as
\be
\rho(\lambda)=\frac{1}{N}\sum_{i=1}^N \moy{\delta(\lambda-\lambda_i)}\;,
\ee
where the $\lambda_i$'s are the random eigenvalues of $W=K T K^T$. In this expression, the expectation is taken over both the random $N\times M$ matrix $K$ with independent $N(0,1/\sqrt{N})$ entries and the random $M\times M$ diagonal matrix $T$ with real i.i.d $N(0,1)$ random elements. The resolvent $m(z)=\int d\lambda\,\rho(\lambda)/(z-\lambda) $, defined as the Stieltjes transform of the spectal density $\rho(\lambda)$, is solution of \eqref{res_eq} with the Gaussian limiting distribution $n(t)=\exp(-t^2/2)/\sqrt{2\pi}$. Computing explicitly the integral over $t$, one obtains the following transcendental equation for $m(z)$,
\begin{equation}\label{MM}
z=\frac{1-\alpha}{m(z)}+\frac{i\alpha}{m^2(z)}\Psi\left(\frac{1}{i\, m(z)}\right)\;,
\end{equation}
where $\Psi(u)=\sqrt{\frac{\pi}{2}}e^{\frac{u^2}{2}}(1+\erf(u/\sqrt{2}))$. Taking the paramter $z=\lambda$ on the real axis, we introduce as in the previous section the real and imaginary part of the resolvent $m(\lambda)=m_r(\lambda)+i m_i(\lambda)$. The spectral density is then extracted from the resolvent $m(\lambda)$ by using the identity
\be
\rho(\lambda)=\frac{d}{d\lambda}\lim_{y\rightarrow 0^+}\int_{\lambda_0}^\lambda \frac{du}{\pi} m_i(u+i y)=\frac{m_i(\lambda)}{\pi}\,.
\ee
While equation \eqref{MM} allows an efficient numerical evaluation of the density at a given real position $z=\lambda$, the properties of the mean density $\rho(\lambda)$ are most easily analysed using the set of equations for the real and imaginary part of the resolvent:
\begin{align}
  \frac{m_r}{m_i^2+m_r^2}&=\lambda-\alpha\int_{-\infty}^{\infty} dt\,\frac{t\,n(t)\,(1-t\,m_r)}{m_i^2\,t^2+(1-t\,m_r)^2}\,,\label{m_r_eq_2}\\
\frac{m_i}{m_i^2+m_r^2}&=\alpha\int_{-\infty}^{\infty} dt\,\frac{t^2\,n(t)\,m_i}{m_i^2\,t^2+(1-t\,m_r)^2}\,.  \label{m_i_eq_2}
\end{align}
In particular, for an even limiting density $n(-t)=n(t)$, one can show from (\ref{m_r_eq_2}-\ref{m_i_eq_2}) that the mean density $\rho(-\lambda)=\rho(\lambda)$ is even as well. Indeed the pair of equations (\ref{m_r_eq_2}-\ref{m_i_eq_2}) is invariant under the symmetry $(\lambda,m_r,m_i)\to (-\lambda,-m_r,m_i)$ together with the change of variable $t\to -t$ in the integrals. As $\rho(\lambda)=m_i(\lambda)/\pi$ the mean density is also even.

For an unbounded limiting density $n(t)$ of $T$, such as the Gaussian density in GMPE, the associated mean spectral density $\rho(\lambda)$ for matrices $W=K T K^T$ is also unbounded \cite{MP67}. In that limit, one expects that $m_i(\lambda)=\pi\rho(\lambda)\propto n(\lambda)\ll m_r(\lambda)$ as $\lambda\to \infty$. Using this approximation and introducing in the integrals of Eqs. \eqref{m_r_eq_2} and \eqref{m_i_eq_2} the change of variable $t=1/m_r+m_i u/m_r^3$, one obtains to leading order in $m_i$
\begin{align}
  \frac{1}{m_r(\lambda)}&\approx \lambda+\frac{\alpha}{m_r^2}\,n\left(\frac{1}{m_r}\right) \int_{-\infty}^{\infty}\frac{u\,du}{u^2+m_r^2}=\lambda\\
  m_i(\lambda)&\approx \alpha\,n\left(\frac{1}{m_r}\right)\int_{-\infty}^{\infty}\frac{m_r\,du}{m_r^2+u^2}\approx \pi\alpha\,n(\lambda)\,.
\end{align}
In particular, for a Gaussian density $n(t)$, it yields the asymptotic behaviour 
\begin{equation}\label{rhoasxgrows}
\rho(\lambda)=\frac{m_i(\lambda)}{\pi}\approx \frac{\alpha}{\sqrt{2\pi}}e^{-\frac{\lambda^2}{2}}\;\;\mbox{ for }|\lambda|\gg 1.  
\end{equation}

For $\alpha>1$, the density $\rho(\lambda=0)=\rho_{\alpha}$ is finite and in the present case it is obtained by setting $z=0$ and $m(z)=i\pi \rho_\alpha$ in Eq. \eqref{MM}, yielding 
\begin{align}
&\rho(\lambda=0)=\rho_\alpha\;,\;\;\alpha>1\;,\;\;\left(1-\frac{1}{\alpha}\right) \sqrt{2\pi}\rho_{\alpha}=e^{\frac{1}{2\pi^2\rho^2_{\alpha}}}\erfc\left(\frac{1}{\sqrt{2}\pi\rho_{\alpha}}\right)\;.
\end{align}
As $\alpha\to 1_+$, one obtains that the value $\rho_\alpha\approx [\sqrt{2\pi}(1-\alpha)]^{-1}$ diverges.

For $\alpha<1$ instead, the mean density displays both a continuous and a discontinuous part. The fraction of zero eigenvalues is given by $1-M/N=1-\alpha$ and is independent of the limiting density $n(t)$ (as long as $T$ does not have a macroscopic number of zero eigenvalues). On the other hand, the continuous part of the density reaches a finite value $\rho_\alpha$ for $\lambda\to 0$. The value of this constant can be obtained by inserting $m(\lambda)=(1-\alpha)/\lambda+\pi \rho_{\alpha}$ in Eq. \eqref{MM} and expanding for $\lambda\to 0$. This yields
\begin{align}
\rho_\alpha&=\lim_{\lambda\to 0}\left[\rho(\lambda)-(1-\alpha)\delta(\lambda)\right]=\frac{\alpha}{(1-\alpha)\sqrt{2\pi}}\;,\;\;\alpha<1\;.    
\end{align}

Finally, in the special case $\alpha=1$, the density is continuous but diverges on approaching the origin as
\be
\rho(\lambda)\approx \frac{1}{(2\pi)^{3/4}\sqrt{|\lambda|}}\;,\;\;\lambda\to 0\;,\;\;\alpha=1\;.
\ee

Let us finally consider the moments $\moy{\lambda^n}=\int_{-\infty}^{\infty} d\lambda\,\lambda^n\,\rho(\lambda)$. As $\rho(\lambda)$ is even, it trivially yields all odd moments vanishing: $\moy{\lambda^{2n+1}}=0$. For any $\alpha$, the even moments can be computed using that for $z\to \infty$,
\begin{align}
m(z)&=\int_{-\infty}^{\infty} d\lambda\,\frac{\rho(\lambda)}{z-\lambda}\approx \frac{1}{z}\sum_{k=0}^{\infty} \int_{-\infty}^{\infty} d\lambda\,\left(\frac{\lambda}{z}\right)^k\rho(\lambda)=\sum_{k=0}^{\infty}\frac{\moy{\lambda^k}}{z^{k+1}}=\sum_{n=0}^{\infty}\frac{\moy{\lambda^{2n}}}{z^{2n+1}}\;.  
\end{align}
Using Eq. \eqref{res_eq}, one obtains that
\begin{align}
z-\frac{1}{m(z)}&=\alpha\int_{-\infty}^{\infty} dt\,\frac{t\,n(t)}{1-t\, m(z)}\approx \alpha\int_{-\infty}^{\infty} dt\,t\,n(t)\sum_{k=0}^{\infty}(t\, m(z))^k\approx\alpha\sum_{n=1}^{\infty} \moy{T^{2n}} m(z)^{2n-1}\;,
\end{align}
where $\moy{T^k}=\int_{-\infty}^{\infty}dt\,t^k\,n(t)$ and we have considered a symmetric distribution $n(-t)=n(t)$. In particular for the Gaussian distribution $\moy{T^{2n}}=2^n\Gamma(n+1/2)/\sqrt{\pi}$ and the lowest moments can be obtained explicitly as
\begin{align}
\moy{\lambda^2}&=\int_{-\infty}^{\infty} d\lambda\,\lambda^2\,\rho(\lambda)=\alpha\;,\\
\moy{\lambda^4}&=\int_{-\infty}^{\infty} d\lambda\,\lambda^4\,\rho(\lambda)=\alpha(3+2\alpha)\;.    
\end{align}

\section{Moments and correlation functions of characteristic polynomials of MPE and GMPE}\label{sec_pol}

In this section we descibe a computation of the $p$-point correlation function of the characteristic polynomial defined in Eq.\eqref{charpol} for  Marchenko-Pastur Ensembles. Such objects are known to provide insights into spectral correlations, and as such are interesting on their own and attract a lot of attention. Note also that the mean total number of equilibria is expressed in Eq.\eqref{N_tot_abs_charpol} in terms of the object not dissimilar to  ${\cal Z}_{\beta,p}({\bm \lambda};T)$ defined in Eq.\eqref{charpol}. However, the presence of the absolute value forced us to resort only to asymptotic analysis for $N\to \infty$ relying on the strong self-averaging hypothesis. In contrast, we will show that the moments which do not contain absolute values can be computed exactly for any finite $N$ without any approximation.

Let us now obtain an explicit expression for ${\cal Z}_{\beta,p}({\bm \lambda};T)$. We first consider the case of a fixed diagonal matrix $T$ with elements that are not necessarily positive. Introduce $p$ pairs of $N$-dimensional vectors $\{\tilde{\bm\psi}_j,\bm\psi_j\}$ with anticommuting/Grassmann components such that $\psi_i\psi_j=-\psi_j\psi_i$, $\int d\psi=0$ and $\int\psi d\psi=1$. Then each of the $p$ determinant factors in the righ-hand side of  Eq.(\ref{charpol}) can be written as the standard Berezin integral $\det(A)=\int D\bm{\psi}D\tilde{\bm{\psi}}\exp\left(\tilde{\bm{\psi}}^T A \bm{\psi}\right)=\int D\bm{\psi}D\tilde{\bm{\psi}}\exp\left(-\Tr( A\tilde{\bm{\psi}}^T\bm{\psi})\right)$ yielding:
\begin{align}
{\cal Z}_{\beta,p}({\bm \lambda};T)&=\moy{\prod_{j=1}^p\det(\lambda_j\mathbb{I}_N-W)}_{K}\label{Z_eq1}=\int D\bm{\psi}D\tilde{\bm{\psi}}\,e^{-\sum_{j=1}^p\lambda_j\Tr\bm{\psi}_j\tilde{\bm{\psi}}^T_j}\prod_{a=1}^M\moy{e^{\frac{T_a}{2}\Tr{Q}^{(s)}\bm{k}_a\bm{k}_a^{T}}}_{{\bm k}_a}\\
&=\int D\bm{\psi}D\tilde{\bm{\psi}}\,e^{-\sum_{j=1}^p\lambda_j\Tr\bm{\psi}_j\tilde{\bm{\psi}}^T_j}\prod_{a=1}^M \det\left(\mathbb{I}_N-\frac{T_a}{N}{Q}^{(s)}\right)^{-\beta/2}\;,\nn
\end{align}
where ${Q}^{(s)}={Q}+{Q}^T$ is the symmetric part of ${Q}=\sum_{j=1}^p\bm{\psi}_j\tilde{\bm{\psi}}^T_j$, and in the second line we have used that $\moy{\cdots}_{{\bm k}_a}$ is the expectation with respect to the Gaussian density $p(\bm k_a)\propto e^{-N\beta \bm k_a^{\dagger}\bm k_a/2}$ for $a^{\rm th}$ column of $K$, which we evaluated  in the third line, Eq.\eqref{Z_eq1}. Introducing the $2p\times 2p$ matrix 
\begin{equation}\label{Qq}
\hat{{Q}}= \begin{bmatrix}
 \tilde{{\bm \psi}}^T_i\bm{\psi}_j& \tilde{\bm{\psi}}_i^T\tilde{\bm{\psi}}_j\\
 -\bm{\psi}_i^T\bm{\psi}_j &  -\bm{\psi}_i^T\tilde{\bm{\psi}}_j
 \end{bmatrix}\;,
\end{equation}
satisfying $\Tr ({Q}^{(s)})^k=-\Tr {\hat Q}^k$, one can show that 
\be
\det\left(\mathbb{I}_N-\frac{T_a}{N}{Q}^{(s)}\right)^{-\beta/2}=\det\left(\mathbb{I}_{(2/\beta)p}-\frac{T_a}{N}{\hat Q}\right)^{\beta/2}\;.
\ee
As the integrand in Eq. \eqref{Z_eq1} only depends on $\hat{Q}$ one can use the so-called "bosonization'' trick proposed in \cite{LSZ08} and replace the integral over anticommuting variables by the invariant integration over unitary matrices $U$ belonging to the group  $C(4/\beta) E(p)$:
\begin{align}\label{eq1U}
&\mathcal{Z}_{\beta,p}(\bm{\lambda};T)=\frac{\int_{C(4/\beta) E(p)}( U^\dagger d U)\det U^{-\frac{\beta N}{2}}e^{\frac{\beta}{2}\Tr{\Lambda U}}\prod_{a=1}^M\Theta_a( U)}{\int_{C(4/\beta) E(p)}( U^\dagger d U)\det U^{-\frac{\beta N}{2}}e^{\frac{\beta}{2}\Tr{ U}}}\;,
\end{align}
where $\Theta_{a,\beta}( U)=\det(\mathbb{I}_{(2/\beta) p}-\frac{T_a}{N} U)^{\beta/2}$ and
\be
\Lambda=\mathbb{I}_{2/\beta}\otimes\begin{bmatrix}
    \lambda_1 & 0 & \dots & 0 \\
    0 &  \lambda_2 & \dots & 0 \\
    \vdots & \vdots & \ddots & \vdots \\
    0 & 0 & \dots &  \lambda_p
  \end{bmatrix}.
\ee
Note that for $\beta=1$, the integration over the group $C4E(p)=CSE(p)$ defines the Circular Symplectic Ensemble, i.e. the ensemble of unitary matrices $U$ with skew symmetric sub-blocks
\be
 U = \begin{bmatrix}
    U_{11} &   U_{12}\\\
  U_{21} &   U_{11}^T
 \end{bmatrix}
\ee
that satisfies 
\be
{P}^{-1} U {P}= U^T\;,\;\;{P}=\begin{bmatrix}
  \mathbb{O}_p & -\mathbb{I}_p\\
\mathbb{I}_p &  \mathbb{O}_p
\end{bmatrix}
\;.
\ee 
On the other hand, for $\beta=2$, the group $C2E(p)=CUE(p)$ gives the standard Circular Unitary Ensemble (CUE).

\subsection{Fixed diagonal matrix $T$}

For a fixed diagonal matrix $T$ we focus our attention on the special case where all $\lambda_j$'s are identical, i.e. ${\bm \lambda}=\lambda{\bm 1}=\lambda(1,\cdots,1)$. In that case, both integrands in the numerator and the denominator of Eq.(\ref{eq1U}) are unitary invariant and hence their ratio can be expressed solely in terms of the integrals over eigenvalues of the unitary matrix $U$. This yields the following expression:
\begin{align}
&\mathcal{Z}_{\beta,p}(\lambda;T)\equiv\mathcal{Z}_{\beta,p}(\lambda{\bm 1}_p;T)=\label{Z_eig}  \frac{\displaystyle 
\int_{[0,2\pi]^p} \prod_{j=1}^p d\theta_{j}\prod_{j=1}^p e^{-i N\theta}h(\lambda;\theta_j)\prod_{k<j}\left|e^{i\theta_j}-e^{i\theta_k}\right|^{\frac{4}{\beta}}}{\displaystyle \int_{[0,2\pi]^p}  \prod_{j=1}^p d\theta_{j}\prod_{j=1}^p e^{-i N\theta+ e^{i\theta_j}}\prod_{k<j}\left|e^{i\theta_j}-e^{i\theta_k}\right|^{\frac{4}{\beta}}}\;, 
\end{align}
where we introduced a function
\be
h(\lambda;\theta)=\exp\left(\lambda e^{i\theta}\right)\prod_{a=1}^M \left(1-\frac{T_a}{N}e^{i\theta}\right)\;.
\ee
For $\beta=2$, using that $\det_{1\leq i,j\leq p}(x_i^{j-1})=\prod_{i<j}^p (x_i-x_j)$ together with the Cauchy-Binet-Andréief formula \cite{And}, Eq.\eqref{Z_eig} can be re-written as the ratio of determinants
\begin{align}\label{determinantal}
\mathcal{Z}_{\beta=2,p}(\mathbf{\lambda};T)&=\moy{\det(\lambda\mathbb{I}-K T K^T)^p}_K=\frac{\det\Big[g_{N,k-j}(\lambda;{ T})\Big]_{j,k=1}^{p} }{\det\Big[q_{N,k-j}\Big]_{j,k=1}^p}\;,
\end{align}
where
\begin{align}
g_{N,m}(\lambda;T)&=\int \frac{d\theta}{2\pi} \,e^{-iN\theta}h(\lambda;\theta)e^{im\theta}=\oint_{|z|=1}\frac{dz}{2i\pi\,z^{N+m+1}}e^{\lambda z}\prod_{a=1}^M\left(1-\frac{T_a}{N}z\right)\;,\\
q_{N,m}&=g_{N,m}(1;\mathbb{O})=\int \frac{d\theta}{2\pi} \,e^{-iN\theta}e^{e^{i\theta}}e^{im\theta}=\oint_{|z|=1}\frac{dz}{2i\pi\,z^{N+m+1}}e^{z}\;.
\end{align}
For each integral here there is a unique pole of order $N+m+1$ at $z=0$, and  using the residue theorem yields
\begin{align}
    g_{N,m}(\lambda;{T})=\lambda^{N+m}\sum\limits_{l=0}^{N+m}\frac{(-1)^l}{(N+m-l)!}e_{l}\left(\frac{T}{N\lambda}\right)\;,\;\;q_{N,m}&=\frac{1}{(N+m)!}\;,
\end{align}
where $e_l(X)$ is the $l$-th elementary symmetric polynomial defined as
\be
e_l(X_1,\cdots,X_n)=\sum_{1\leq i_1<i_2<\cdots<i_l\leq n}\prod_{j=1}^l X_{i_j}\;.
\ee

Similarly, for $\beta=1$ using the identity 
\begin{align}
\prod_{j<k}|e^{i\theta_j}-e^{i\theta_k}|^4=\prod_{j=1}^p e^{-2i(p-1)\theta_j}\det_{1\le k\le 2p,1\le j\le p}\Big[e^{i(k-1)\theta_j}; (k-1)e^{i(k-2)\theta_j} \Big]\;,    
\end{align}
and applying further the De Bruijn's identity, the $p$-th moment of the characteristic polynomial can be expressed as a ratio of Pfaffians:
\begin{align}\label{Pfaffian}
\mathcal{Z}_{\beta=1,p}(\mathbf{\lambda};T)&=\moy{\det(\lambda\mathbb{I}-K T K^T)^p}_K=
\frac{\Pf\Big[(j-k)g_{N,2p+1-(k+j)}(\lambda;{ T})\Big]_{j,k=1}^{2p}}{\Pf\Big[(j-k)q_{N,2p+1-(k+j)}\Big]_{j,k=1}^{2p}}\;.
\end{align}

\begin{figure}[t!]
\centering
 {\includegraphics[width=.49\textwidth]{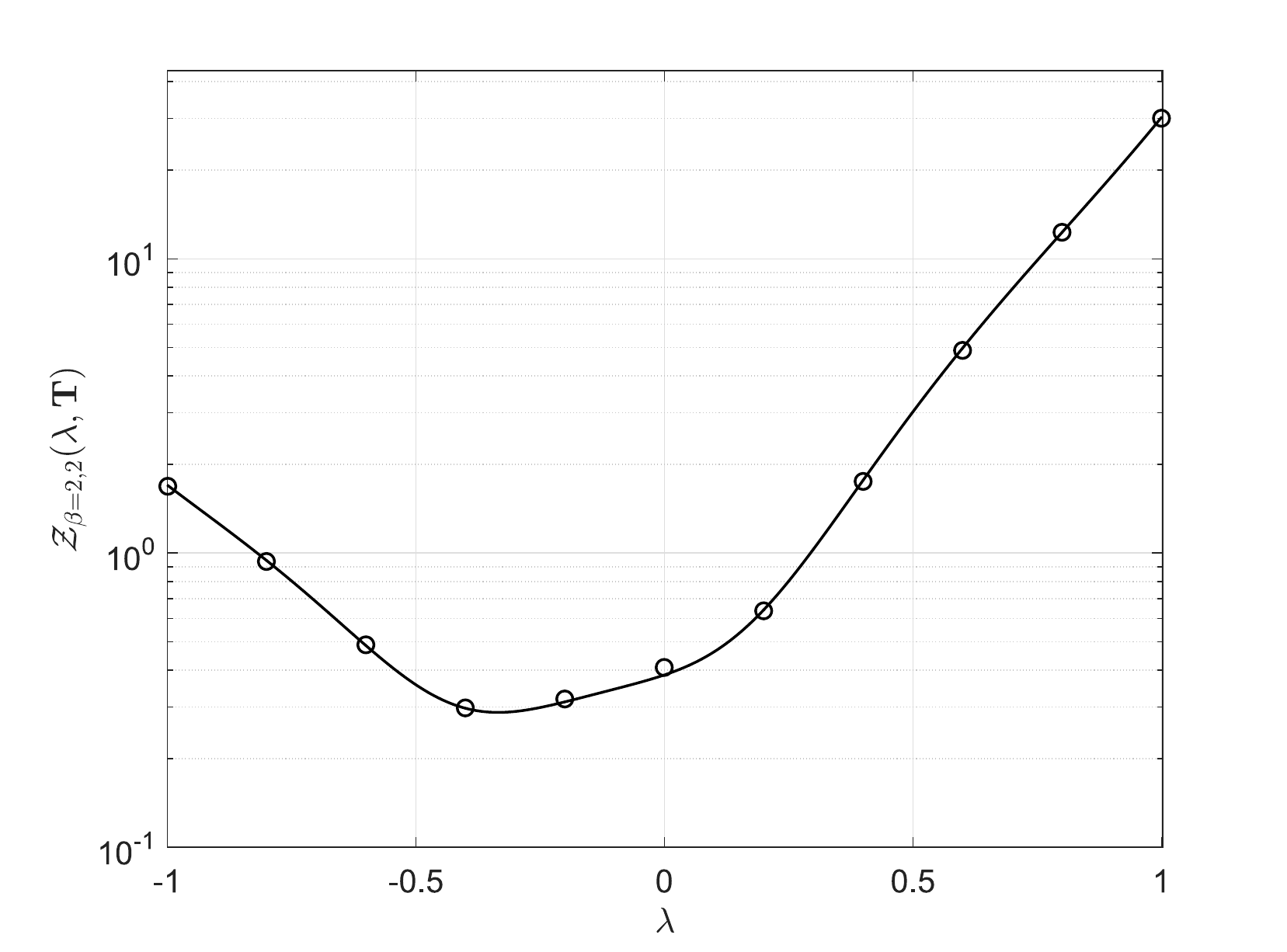}} 
  {\includegraphics[width=.49\textwidth]{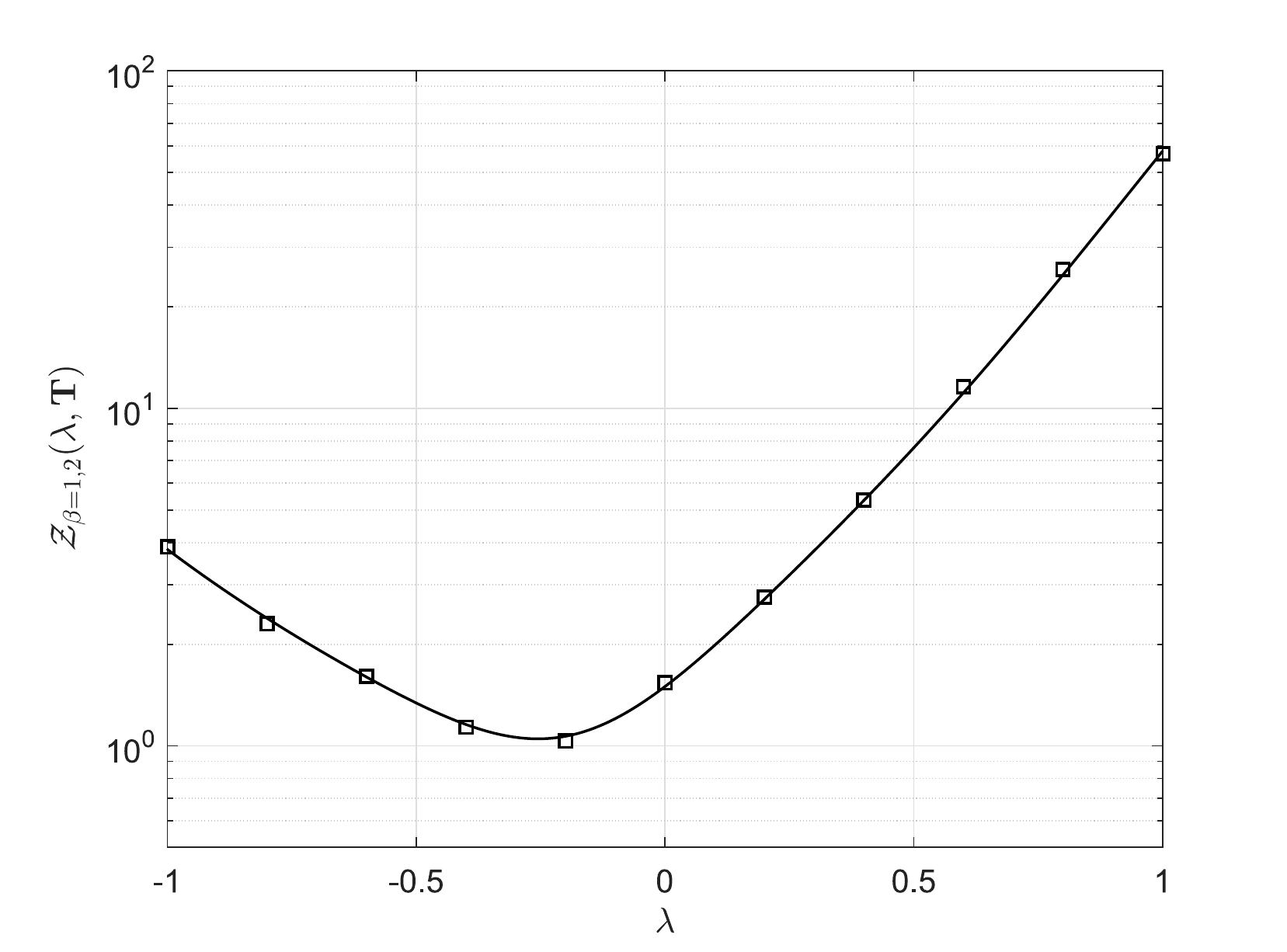}} 
\caption[Plot of $\mathcal{Z}_{2,2}(\lambda,T)$
and $\mathcal{Z}_{1,2}(\lambda,T)$ for fixed $T$
against their numerical simulations]{Left: Comparison between the analytical expression for  $\mathcal{Z}_{2,2}(\lambda,T)$ in Eq.(\ref{determinantal}) (dark line) for $N=5$, $\alpha=M/N=2$ and fixed $T^{(1)}=(0.4339,1.0562,-0.3710,0.3090,-0.8655,$ $-1.0482,-1.6036,-0.7595,1.1370,-1.0643)^T$ and numerical simulations (markers), plotted as a function of $\lambda$. Right: Comparison between the analytical expression for  $\mathcal{Z}_{1,2}(\lambda,T)$ in Eq.(\ref{Pfaffian}) (dark line) for $N=4$, $\alpha=M/N=2$ and fixed $T^{(2)}=(0.4339,1.0562,-0.3710,0.3090,$ $-0.8655,-1.0482,-1.6036,-0.7595)^T$ and numerical simulations (markers), plotted as a function of $\lambda$. For each figure the number of samples is $10^5$.}
\label{detFixed}
\end{figure}

In Fig. \ref{detFixed}, we have compared our exact analytical results for $\mathcal{Z}_{\beta,p}(\mathbf{\lambda};T)$ as a function of $\lambda$  given in Eq. \eqref{Pfaffian} and \eqref{determinantal} respectively for $\beta=1,2$ taking fixed diagonal matrix $T$ with numerical simulations, obtaining an excellent agreement.

Let us finally mention that in the case $\lambda=0$ one can use the following identities from the book \cite{Fbook} 
\begin{align}
&\int_{[0,2\pi]^p} \prod_{j=1}^p \left(\frac{d\theta_{j}\,e^{-iN\theta_j}}{2\pi}e^{e^{i\theta_j}} \right)\prod_{j<k}|e^{i\theta_k}-e^{i\theta_j}|^{\frac{4}{\beta}}=\\
&M_{p}\left(-\frac{N}{2},\frac{N}{2},\frac{2}{\beta}\right){}_{1}F_{1}^{\left(\frac{\beta}{2}\right)}\left(-\frac{N}{2},-\frac{N}{2}+1+\frac{2}{\beta}(p-1);-1\right)\;,\nn\\
&\int_{[0,2\pi]^p} \prod_{j=1}^p \left(\frac{d\theta_{j}\,e^{-iN\theta_{j}}}{2\pi} \prod_{a=1}^M\left(1-\frac{T_a}{N} e^{i\theta_j}\right)\right)\prod_{j<k}|e^{i\theta_k}-e^{i\theta_j}|^{\frac{4}{\beta}}=\\
&M_{p}\left(-\frac{N}{2},\frac{N}{2},\frac{2}{\beta}\right){}_{2}F_{1}^{\left(\frac{2}{\beta}\right)}\left(-p,\frac{\beta N}{4};(1-p)+\frac{\beta}{2}\left(\frac{N}{2}-1\right);-\frac{T}{N}\right)\;,\nn
\end{align}
with
\be
    M_p(a,b,c)=\prod_{j=0}^{p-1}\frac{\Gamma(cj+a+b+1)\Gamma(c(j+1)+1)}{\Gamma(cj+a+1)\Gamma(cj+b+1)\Gamma(1+c)}
\ee
and where the hypergeometric function of matrix argument is defined as
\be
_pF_q^{(\alpha)}({\bf a};{\bf b};X)=\sum_{k=0}\sum_{\kappa\vdash k}\frac{\prod_{j=1}^p(a_j)^{(\alpha)}_\kappa}{\prod_{j=1}^{q}(b_j)^{(\alpha)}_\kappa}\frac{C_{\kappa}^{(\alpha)}(X)^k}{k!}\;,
\ee
with $\kappa$ standing for a partition of $k$, $(a)^{(\alpha)}_\kappa$ being the generalised Pochhammer symbol associated to the partition $\kappa=(\kappa_1,\cdots,\kappa_m)$,
\be
(a)^{(\alpha)}_{\kappa}=\prod_{i=1}^{m}\prod_{j=1}^{\kappa_i}\left(a-\frac{i-1}{\alpha}+j-1\right)\;,
\ee
and the function $C_{\kappa}^{(\alpha)}(X)$ is the so-called $C$-normalisation of Jack polynomial \cite{Fbook}.
This allows to express the function $\mathcal{Z}_{\beta,p}(0;T)$ via a ratio of the hypergeometric functions as
\begin{align}
\mathcal{Z}_{\beta,p}(0;T)&=\moy{\det(-W)^p}_K=\frac{{}_{2}F_{1}^{\left(\frac{2}{\beta}\right)}\left(-p,\frac{\beta N}{4};(1-p)+\frac{\beta}{2}\left(\frac{N}{2}-1\right);-\frac{T}{N}\right)}{{}_{1}F_{1}^{\left(\frac{\beta}{2}\right)}\left(-\frac{N}{2},-\frac{N}{2}+1+\frac{2}{\beta}(p-1);-1\right)}\;.
\end{align}

\subsection{Random diagonal matrix $T$}

Let us now consider the case of random diagonal matrix $T$ and average the $p$-point characteristic polynomial $\mathcal{Z}_{\beta,p}({\bm \lambda};T)$ over realisations of $T$, denoting these expectations as $\mathcal{Z}_{\beta,p}({\bm \lambda})=\moy{\mathcal{Z}_{\beta,p}({\bm \lambda};T)}_T$. In contrast to the case of fixed $T$, we now consider $\lambda_j$'s that are not necessarily identical but focus only on the low values $p=1$ and $p=2$ for $\beta=1,2$. Interestingly, we find that the ensuing formulae are universal in the sense that $\mathcal{Z}_{\beta,p}({\bm \lambda})$ depends on the whole distribution of $T$ only via the lowest moments $\moy{T^k}=\int_{-\infty}^{\infty} dt\,t^k\,n(t), \, k=1,2$ of the i.i.d. elements of $T$, assuming those moments are finite.

\subsubsection{The case $p=1$}

\begin{figure}[t!]
\centering
 \includegraphics[width=.8\textwidth]{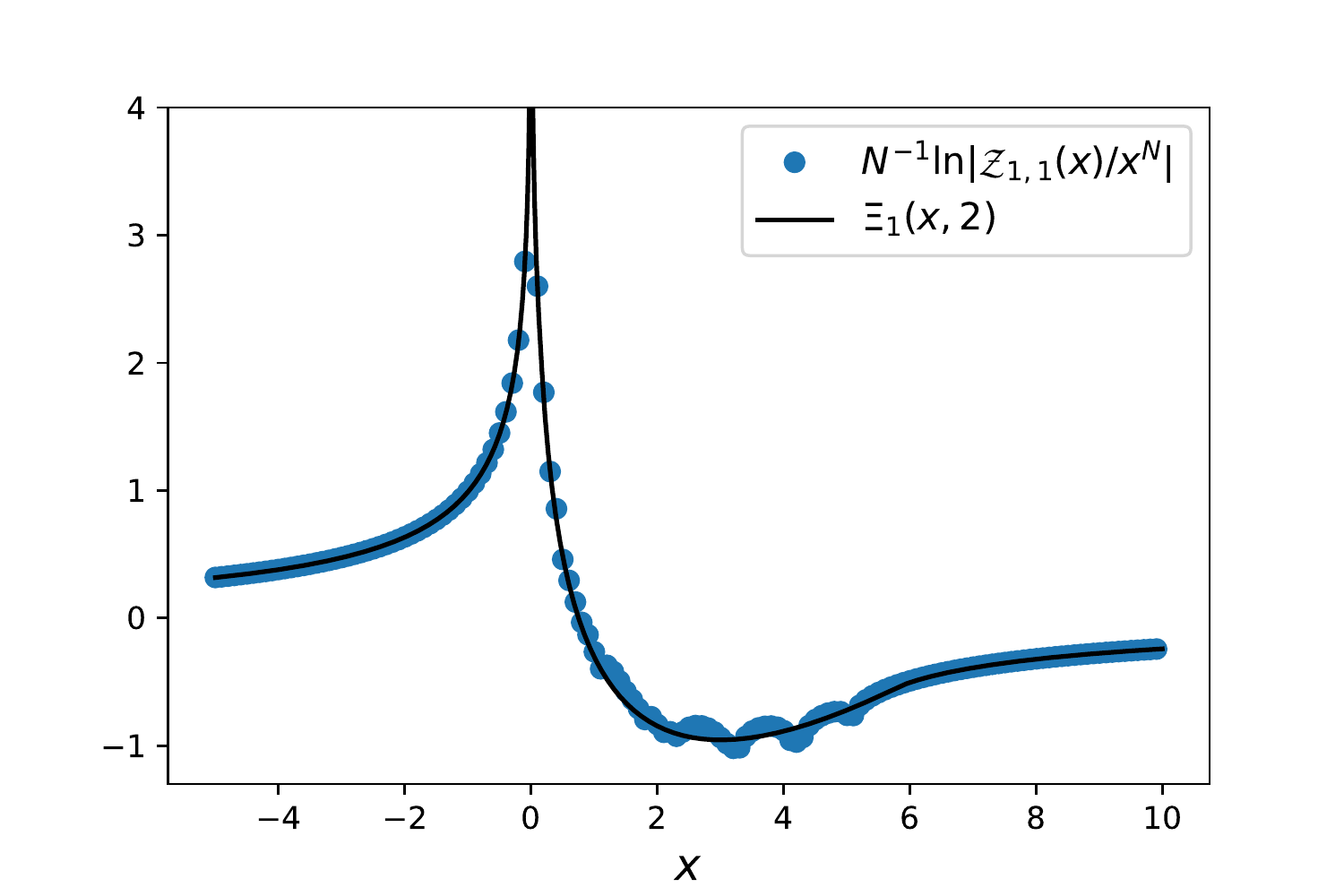}
\caption{Comparison between the analytical expression of $\Xi_1(x,\alpha)$ for $\alpha=2$ (black solid line) obtained from Eq. \eqref{Xi_1_an} and the expression of $\frac{1}{N}\ln|{\cal Z}_{1,1}(x)/x^N|$ (blue circles) for $\moy{T}=1$, $N=20$ and $\alpha=2$, plotted as a function of $x$. The number of samples is $10^7$.}
\label{Xi_1_fig}
\end{figure}

In such a case we can directly use the results in Eq. \eqref{determinantal} and \eqref{Pfaffian} for fixed $T$ and take the average value with respect to the i.i.d. diagonal elements $T_a$ of $T$. This yields 
\begin{align}
\mathcal{Z}_{\beta,1}(\lambda)&=\moy{\mathcal{Z}_{\beta,1}(\lambda;T)}_T\label{Z_b_1}=N!\displaystyle \oint_{|z|=1} \frac{dz}{2\pi\,i\,z^{N+1}}e^{\lambda z}\left(1-\frac{\moy{T}}{N}z\right)^M\;.  
\end{align}
In the particular case of a symmetric distribution $n(t)=n(-t)$ we have $\moy{T}=0$ and \eqref{Z_b_1} simplifies drastically: $\mathcal{Z}_{\beta,1}(\lambda)=\moy{\det(\lambda\mathbb{I}-W)}=\lambda^{N}$. If $\moy{T}\neq 0$ instead and assuming $|\moy{T}|<N$, $\mathcal{Z}_{\beta,1}(\lambda)$ is expressed for finite $N$ as
\begin{align}\label{eqZ1}
&\mathcal{Z}_{\beta,1}(\lambda)=\left(\frac{\moy{T}}{N}\right)^N\frac{M!}{(M-N)!}{}_1F_1\left(-N;1+M-N;\frac{\lambda N}{\moy{T}}\right)\;.
\end{align}
where we remind the definition of the standard hypergeometric funtion
\be
_pF_q({\bf a};{\bf b};x)=\sum_{k=0}\frac{\prod_{j=1}^p(a_j)_k}{\prod_{j=1}^{q}(b_j)_k}\frac{x^k}{k!}\;,
\ee
with $(a)_n=\prod_{i=0}^{n-1}(a+i)$ the Pochhammer symbol.

Note that the ratio $\mathcal{Z}_{\beta,1}(\lambda)/\lambda^N$ only depends on the rescaled variable $x=\lambda/\moy{T}$. In the limit $N\to \infty$ and for $\moy{T}\neq 0$, changing variable from $z$ to $w=(\moy{T}/N)z$ in Eq.\eqref{Z_b_1}, one obtains 
\begin{align}
&\frac{\mathcal{Z}_{\beta,1}(\lambda)}{\lambda^N}=\frac{N!}{N^Ne^{-N}}\displaystyle \oint \frac{dw}{2\pi\,i\,w}e^{ N{\cal L}_1\left(w;x,\alpha\right)}\;,\;\;x=\frac{\lambda}{\moy{T}}\;,\label{Z_1_b_sp}\\
&{\cal L}_1(w;x,\alpha)=x w-\ln x w+\alpha\ln\left(1-w\right)-1
\end{align}
which can then be conveniently analysed using Laplace's method. The saddle points of ${\cal L}_1(w;x,\alpha)$ in the $w$ domain are given by $w_{\pm}(x,\alpha)=(1-\alpha+x\pm\sqrt{(\alpha-1-x)^2-4x})/(2x)$ and are complex conjugate for $(\alpha-1-x)^2<4x$, i.e. $x\in [x_-(\alpha),x_+(\alpha)]$ where $x_{\pm}(\alpha)=(1\pm\sqrt{\alpha})^2$ or both real in the converse case. Interestingly, the boundaries of the interval where $w_{\pm}$ are complex conjugate correspond to the boundaries of the support of the classic Marchenko-Pastur distribution, i.e. $x_\pm(\alpha)=(1\pm\sqrt{\alpha})^2$. In the large $N$ limit, a careful analysis of Eq.\eqref{Z_1_b_sp} (see Appendix \ref{app_p_1} for details) allows to obtain
\begin{align}
\Xi_{1}(x;\alpha)&=\lim_{N\to \infty}\frac{1}{N}\ln\left|\frac{\mathcal{Z}_{\beta,1}(\lambda)}{\lambda^N}\right|\label{Xi_1_an}=\begin{cases}
{\cal L}_1(w_+;x,\alpha)&\;,\;\;x<x_-(\alpha)\;, \\
&\\
\Re\left[{\cal L}_1(w_\pm;x,\alpha)\right]&\;,\;\;x_-(\alpha)\leq x\leq x_+(\alpha)\\
&\\
{\cal L}_1(w_-;x,\alpha)&\;,\;\;x>x_+(\alpha)\;.
\end{cases}  
\end{align}
The function $\Xi_{1}(x;\alpha)$ changes sign over the real interval and behaves in the limit $x\to \pm \infty$ as
\be
\Xi_{1}(x;\alpha)\approx -\frac{\alpha }{x}\;,\;\;x=\frac{\lambda}{\moy{T}}\to \pm \infty\;,
\ee
which matches smoothly the expression $\ln |\mathcal{Z}_{\beta,1}(\lambda)/\lambda^N|=0$ for $\moy{T}=0$. This result is in stark contrast with the behaviour of $\Xi_{\rm tot}(\mu;\alpha)$, which is positive for any value of $\mu>0$ and vanishes as $\mu\to \infty$ as in Eq.\eqref{xi_tot_mu_large}, i.e. much faster than the algebraic decay, emphasising the crucial role played by the absolute value in the definition \eqref{xi_tot_abs_det}. In Fig.\ref{Xi_1_fig}, we show a comparison between the analytical expression of $\Xi_{1}(x,\alpha)$  for $\alpha=2$ and the finite $N=20$ expression of $N^{-1}\ln|{\cal Z}_{1,1}(x)/x^N|$ with $\moy{T}=1$, showing an excellent agreement.

\subsubsection{Calculation for $p=2$ and $\beta=1,2$}

\begin{figure}[t!]
\centering
 {\includegraphics[width=.49\textwidth]{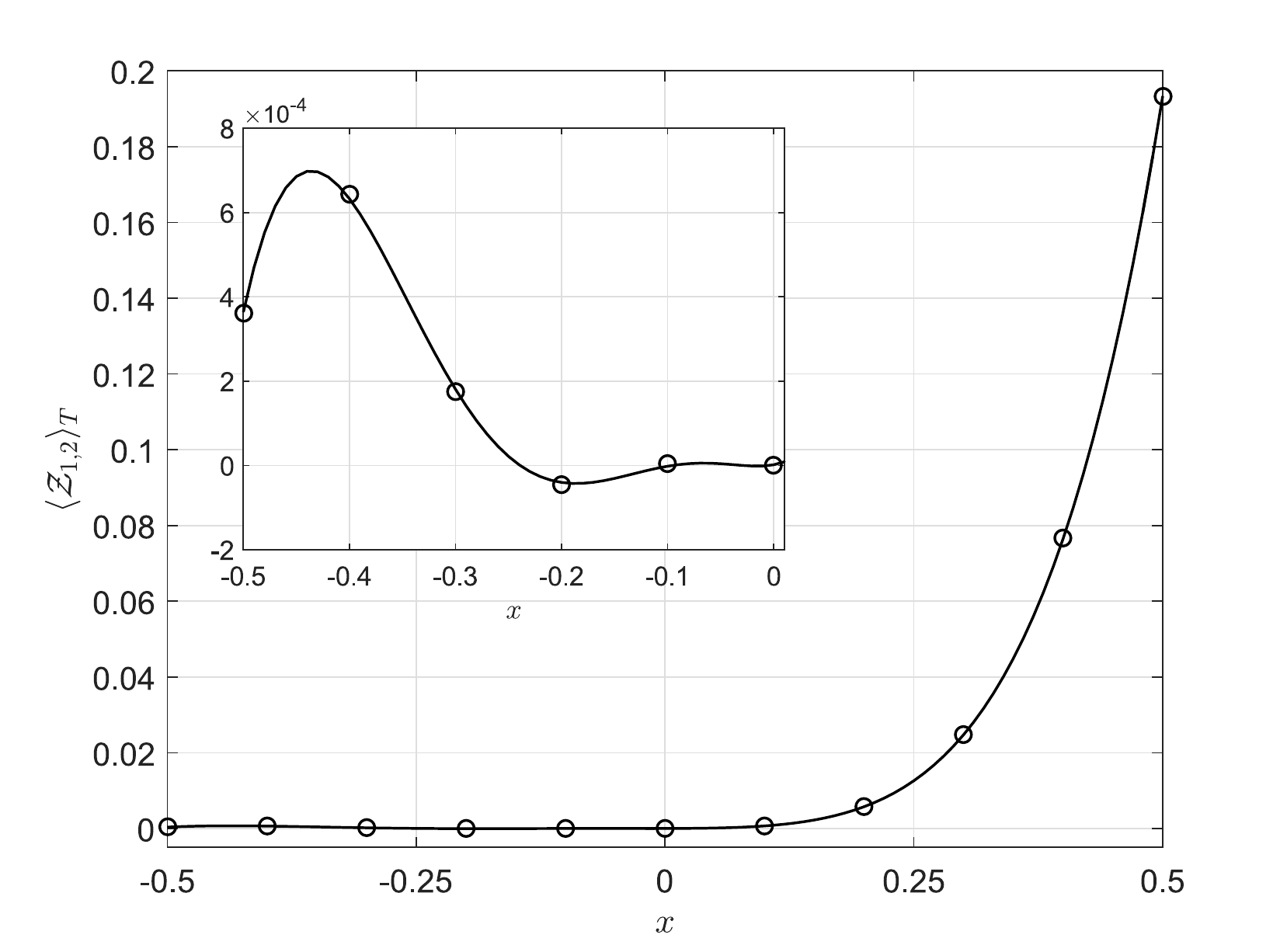}} 
 {\includegraphics[width=.49\textwidth]{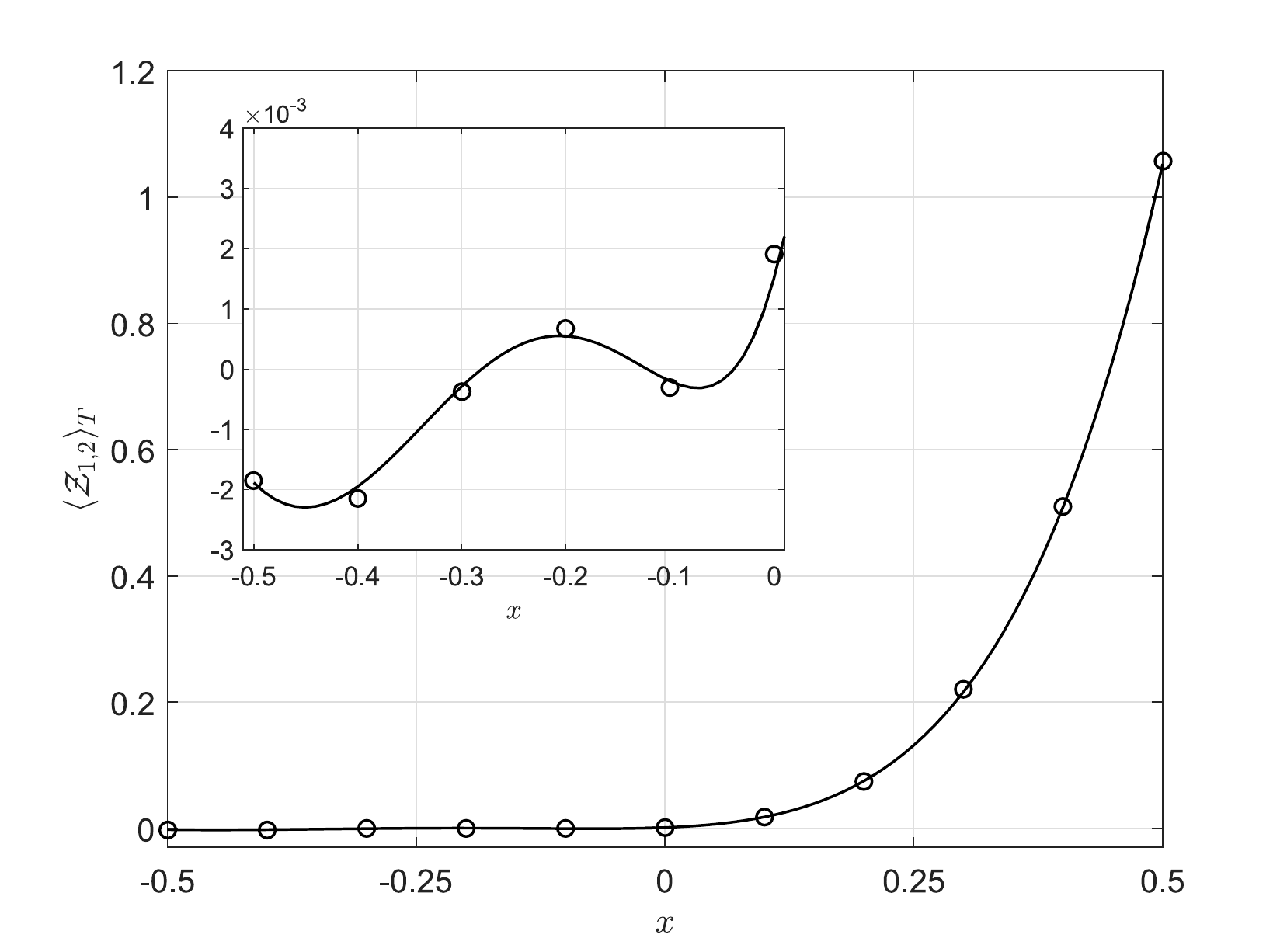}} 
\caption{Comparison between the analytical expression of $\moy{\mathcal{Z}_{1,2}}_T$ in Eq.(\ref{eq3}) (solid line) and numerical simulations (markers), plotted as a function of the product of spectral parameters $x=\lambda_1\lambda_2$. The left figure corresponds to $\alpha=1$ while the right one corresponds to $\alpha=2$. For each figure we have fixed $\moy{T^2}=0.5,\moy{T}=0,N=6$ and the number of samples is $10^6$.}
\label{eqp2}
\end{figure} 

In order to obtain an exact finite $N$ expression of the $2$-point correlation function ${\cal Z}_{\beta,2}({\bm \lambda})=\moy{{\cal Z}_{\beta,2}({\bm \lambda};T)}_T$ for ${\bm \lambda}=(\lambda_1,\lambda_2)$, we start from Eq.\eqref{eq1U} and introduce an explicit representation of matrices $U\in C(4/\beta)E(2)$, separately for $\beta=1$ and $\beta=2$. In the special case of vanishing moment $\moy{T}=0$ one is able to perform the integration over matrices $U$ (see the Appendix \ref{app_Z_2_b_N_fin} for details), and obtain the following expression:
\begin{align}\label{eq3}
&\mathcal{Z}_{\beta,2}(\bm{\lambda}=(\lambda_1,\lambda_2))=\\
&\bigg(\frac{\moy{T^2}}{N^2}\bigg)^N\frac{\beta}{2}\Gamma\left(N+\frac{2}{\beta}+1\right)\frac{M!}{(M-N)!}{}_1F_2\bigg(-N;\frac{2}{\beta}+1,1+M-N;-\frac{ \lambda_1\lambda_2 N^2}{\moy{T^2}}\bigg)\;.\nn
\end{align}
Note that  $\mathcal{Z}_{\beta,2}(\bm{\lambda})$ only depends on ${\bm \lambda}=(\lambda_1,\lambda_2)$ through the product $\lambda_1\lambda_2$, and  the ratio $\mathcal{Z}_{\beta,2}(\bm{\lambda}=(\lambda_1,\lambda_2))/(\lambda_1\lambda_2)^{N}$ only depends on the rescaled variable $y=\lambda_1\lambda_2/\moy{T^2}$.
This is highly unusual as similar objects for Gaussian or Wishart matrices   become in the large-$N$ limit a function of the difference $|\lambda_1-\lambda_2|$, see \cite{BH1} and \cite{ABDFbook}. In Fig. \ref{eqp2}, we show a comparison between our analytical result in Eq. \eqref{eq3} and numerical simulations, showing excellent agreement.

\begin{figure}[t!]
\centering
 \includegraphics[width=.8\textwidth]{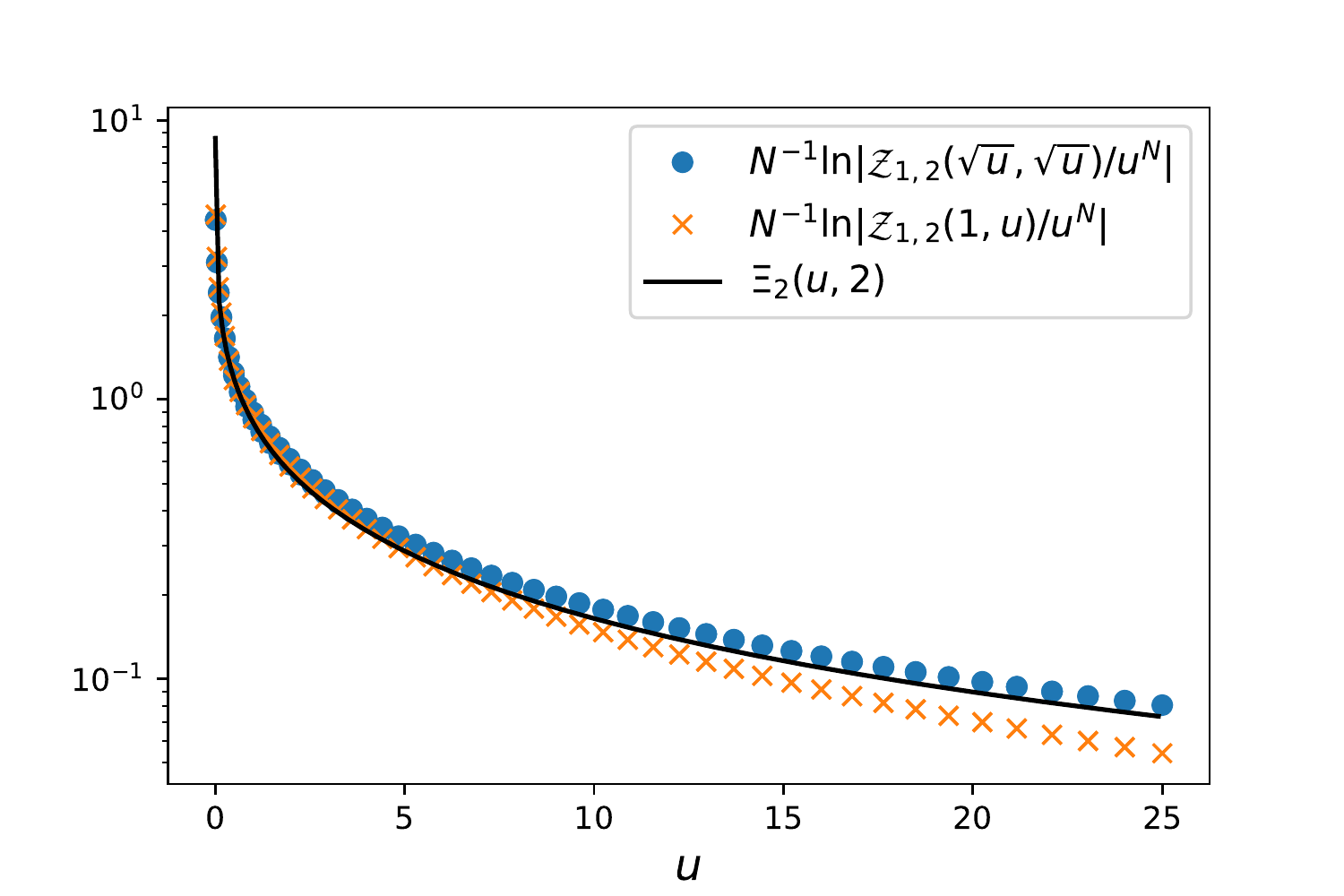}
\caption{Comparison between the analytical expression of $\Xi_2(x,\alpha)$ for $\alpha=2$ (black solid line) obtained from Eq. \eqref{Xi_2_an} and the expression of $N^{-1}\ln|{\cal Z}_{1,2}(\sqrt{u},\sqrt{u})/u^N|$ (blue circles) and $N^{-1}\ln|{\cal Z}_{1,2}(1,u)/u^N|$ (orange triangles) for $\moy{T}=0$, $\moy{T^2}=0$, $N=20$ and $\alpha=2$, plotted as a function of $x$. The number of samples is $10^7$. It is expected that all three curves coincide in the limit $N\to \infty$. The agreement is rather good for low values of $u$ but there is some discrepancy for larger values of $u$, most probably due to finite $N$ effects.}
\label{Xi_2_fig}
\end{figure} 

Let us consider the large $N$ limit of Eq. \eqref{eq3}. To this purpose, we re-write the ratio of Eq.\eqref{eq3} by $(\lambda_1\lambda_2)^N$ as the finite sum
\begin{align}
   &\frac{\mathcal{Z}_{\beta,2}(\bm{\lambda})}{(\lambda_1\lambda_2)^N}=\sum_{k=0}^N\frac{\beta}{2 (N^2 y)^{N}}\frac{ (N-k+1)_k  \Gamma (M+1) \Gamma (N+\frac{2 }{\beta}+1)}{\left(\frac{2}{\beta}+1\right)_k \Gamma (M -N+1) (M -N+1)_k}\frac{\left(N^2 y\right)^k}{k!}\,,
\end{align}
where $y=\lambda_1\lambda_2/\moy{T^2}$. Replacing $M=\alpha N$ with $\alpha\geq 1$, and transforming the sum over $k$ into a Riemann integral over $\eta=k/N$, one can then evaluate the large $N$ behaviour of this expression using the Laplace method, such that
\begin{align}
\Xi_{2}(u;\alpha)&=\lim_{N\to \infty}\frac{1}{N}\ln\left|\frac{\mathcal{Z}_{\beta,2}({\bm \lambda})}{(\lambda_1\lambda_2)^N}\right|=\max_{\eta\in[0,1]}{\cal L}_2(\eta;u,\alpha)\;,\;\;u=|y|=\frac{|\lambda_1\lambda_2|}{\moy{T^2}}\;,
\end{align}
where (see Appendix \ref{app_LD_Z_b_2} for details)
\begin{align}
{\cal L}_2(\eta;u,\alpha)=&\alpha\ln(\alpha)-(\alpha+\eta-1)\ln (\alpha+\eta-1)\\
&+\eta\ln (u(1-\eta ))+2\eta-2\eta\ln(\eta)-\ln (1-\eta)-2-\ln u\nn\;.
\end{align}
The saddle-point equation yields a third order algebraic equation for the value $\eta$ that maximises the action ${\cal L}_2(\eta;u,\alpha)$
\begin{align}
&\left.\partial_{\eta}{\cal L}_2\right|_{\eta=\eta_*}= -\ln(\alpha+\eta_*-1)+\ln(1-\eta_*)+\ln(u(1-\eta_*))-2\ln \eta_*=0\\
&\Rightarrow \eta_*=\frac{1-\alpha}{3}+r_*\;\;{\rm with}\;\;-\frac{\Delta}{27}+\frac{\Theta}{3} r_*+r_*^3=0\;,\\
&{\rm where}\;\;\Delta=9u(2+\alpha)-2(\alpha-1)^3\;,\;\;\Theta=3 u-(\alpha -1)^2\;,
\end{align}
and where the correct solution is obtained using Cardano's formula as (see Appendix \ref{app_LD_Z_b_2})
\be
\eta_*=\frac{1-\alpha }{3}-\frac{\sqrt[3]{2} \Theta}{3 \sqrt[3]{\Delta +\sqrt{\Delta ^2+4 \Theta^3}}}+\frac{\sqrt[3]{\Delta+\sqrt{\Delta ^2+4 \Theta^3}}}{3 \sqrt[3]{2}}\;.
\ee
Thus, for any value of $u$ and $\alpha$, one can show that
\begin{align}
\Xi_{2}(u;\alpha)&=\lim_{N\to \infty}\frac{1}{N}\ln\left|\frac{\mathcal{Z}_{\beta,2}({\bm \lambda})}{(\lambda_1\lambda_2)^N}\right|={\cal L}_2(\eta_*;u,\alpha)\;,\;\;u=\frac{|\lambda_1\lambda_2|}{\moy{T^2}}\;.\label{Xi_2_an}   
\end{align}
Similarly as for the total complexity $\Xi_{\rm tot}(\mu;\alpha)$, the function $\Xi_{2}(u;\alpha)$ is positive for any finite $u$ and vanishes only as $u\to \infty$ although in a much slower fashion compared to Eq.\eqref{xi_tot_mu_large}, i.e.
\be
\Xi_{2}(u;\alpha)\approx \frac{\alpha}{u}\;,\;\;u\to \infty\;.
\ee
On the other hand, one can obtain that the function $\mathcal{Z}_{\beta,2}({\bm \lambda})$ achieves a value exponentially large in $N$ for $\lambda_1\lambda_2=0$, with
\be
\lim_{N\to \infty}\frac{1}{N}\ln\mathcal{Z}_{\beta,2}({\bm \lambda})=\alpha\ln \alpha+(\alpha-1)\ln(\alpha-1)-2\;.
\ee
In Fig. \ref{Xi_2_fig}, we show a comparison between the expression of $\Xi_{2}(u;\alpha)$ in Eq. \eqref{Xi_2_an} and numerical simulations of $N^{-1}\ln|{\cal Z}_{1,2}(\sqrt{u},\sqrt{u})/u^N|$ and $N^{-1}\ln|{\cal Z}_{1,2}(1,u)/u^N|$. In the large $N$ limit, it is expected that $\lim_{N\to \infty} N^{-1}\ln|{\cal Z}_{1,2}(\lambda_1,\lambda_2)/(\lambda_1\lambda_2)^N|$ only depends on the rescaled parameter $u=|\lambda_1\lambda_2|$ and these quantities should thus coincide. The agreement is reasonably good for low values of $u$ but there is some discrepancy (probably resulting from finite $N$ corrections) for larger values of $u$.

\section{Conclusion}\label{sec_conclu}

In this article, we have introduced a model of complex landscape resulting from a superposition of a large number $M$ of random plane waves (larger than the dimension $N$ of the Euclidean space). We have computed exactly the total annealed complexity of this random landscape confined by an isotropic harmonic potential of strength $\mu$ in the limit $N,M\to \infty$ with $\alpha=M/N>1$. In contrast to previously studied cases, see e.g. \cite{F04}, this system does not display a "topological trivialisation transition", i.e. the total annealed complexity is strictly positive for any finite $\mu$. 

The random part of the Jacobian associated to such a random landscape defines a random matrix $W=K T K^T$ belonging to a new random matrix ensemble that we denote "Gaussian Marchencko Pastur Ensemble" (GMPE). We have studied the properties of this ensemble by computing both the average density of eigenvalues in the large $N$ limit and some specific correlation functions of the characteristic polynomial. Rather surprisingly, we have shown that in the large $N$ limit, the two-point correlation function depends on the product of the two spectral parameters instead of their distance, as is the case for standard random matrix ensembles \cite{BH1,ABDFbook}.

The unforeseen behaviour of the total annealed complexity naturally invites further study of the Gibbs measure associated to this random landscape at fixed inverse temperature  $\beta=1/T$. Indeed, for translationally and rotationally invariant Gaussian models \cite{FW07}, it was shown that the trivial phase naturally corresponds to the replica symmetric phase in the Parisi framework. The absence of such phase in the complexity might indicate that the replica symmetry breaking is operative for any $\mu$ at $T=0$ and possibly also for positive values of the inverse temperature in this system. This will be the subject of a future publication.

Finally, it would be worth performing some numerical studies on the number of stationary points, their type, as well as on other landscape statistics for the type of landscapes studied in the present paper. We hope that for moderately big values of $N$ such simulations should be more feasible than for landscapes based on more standard rotationally and translationally invariant Gaussian models.

\newpage

\appendix

\section{Laplace approximation for ${\cal Z}_{\beta,1}$}\label{app_p_1}

Let us start from Eq.\eqref{Z_1_b_sp} that we reproduce here
\begin{align}
\frac{\mathcal{Z}_{\beta,1}(\lambda)}{\lambda^N}&=\frac{N!}{N^Ne^{-N}}\displaystyle \oint \frac{dw}{2\pi\,i\,w}e^{ N{\cal L}_1\left(w;x,\alpha\right)}\;,\\
{\cal L}_1(w;x,\alpha)&=x w-\ln x w+\alpha\ln\left(1-w\right)-1\;,
\end{align}
where $x=\lambda/\moy{T}$. The saddle points of ${\cal L}_1(w;x,\alpha)$ are given by 
\be
w_{\pm}=\frac{1-\alpha+x\pm\sqrt{(\alpha-1-x)^2-4x}}{2x}\;.
\ee
The contribution of each of them depends on the sign of $\Delta_1=(\alpha-1-x)^2-4x$. 

For $\Delta_1>0$,  corresponding either to $x<x_-(\alpha)$ or $x>x_+(\alpha)$ with $x_\pm(\alpha)=(1\pm \sqrt{\alpha})^2$, the saddle points $w_{\pm}$ both lie on the real line. 
In order to obtain the stability of the fixed points, let us compute
\be
\left.\partial_{w}^2{\cal L}_1(w;x,\alpha)\right|_{w=w_\pm}=\frac{1}{w_{\pm}^2}-\frac{\alpha}{(1-w_{\pm})^2}\;,\label{L_1_d_2}
\ee
which is real and positive only for $w_{\pm}$ real and inside the interval between its two roots $((1-\sqrt{\alpha})^{-1},(1+\sqrt{\alpha})^{-1})$. These two roots translate in terms of $x$ as the two edges $x_{\pm}(\alpha)$ of the usual Marchencko Pastur distribution and one can check that for $x>x_+(\alpha)$, one has $w_-\in((1-\sqrt{\alpha})^{-1},(1+\sqrt{\alpha})^{-1})$ (and $w_+$ outside of this interval) and conversely for $x<x_-(\alpha)$, one has $w_+\in((1-\sqrt{\alpha})^{-1},(1+\sqrt{\alpha})^{-1})$ (and $w_-$ outside).

Conversely, for $\Delta_1<0$, i.e. $x\in(x_-(\alpha),x_+(\alpha))$,  $w_{\pm}$ are complex conjugate complex numbers and both contribute in encircling the origin. In this region, the real part of the term in Eq.\eqref{L_1_d_2} is positive and the same for both $w_\pm$, such that both saddle-points have equal contribution. The imaginary part of the term in Eq.\eqref{L_1_d_2} changes sign for $x=(\alpha-1)^2/(\alpha+1)$.

Summarizing, one obtains for $N\gg1$
\begin{align}
&\frac{\mathcal{Z}_{\beta,1}(\lambda)}{\lambda^N}=\begin{cases}
I(w_+) &\;,\;\;x<x_-(\alpha) \\
-I(w_+)-I(w_{-}) &\;,\;\;x_-(\alpha)<x<\frac{(\alpha-1)^2}{\alpha+1} \\
I(w_+)+I(w_{-}) &\;,\;\;\frac{(\alpha-1)^2}{\alpha+1}<x<x_+(\alpha) \\
I(w_{-}) &\;,\;\;x>x_+(\alpha)\\
\end{cases}\\
    &{\rm where}\;I(w_\pm)=\frac{e^{N{\cal L}_1(w_\pm;x,\alpha)}}{\sqrt{2\pi N w_\pm^2\partial_w^2{\cal L}_1(w_\pm;x,\alpha)}}
\end{align}
Finally, for $x=x_\pm(\alpha)$, the saddle points are unique and of order 2:\\
1) For $x=x_+(\alpha)$ one has that $w_{+}=(1+\sqrt{\alpha})^{-1}$, such that
\begin{align}
\left.\partial_{w}^2{\cal L}_1(w;x,\alpha)\right|_{w=w_+}=0\;,\;\;\left.\partial_{w}^3{\cal L}_1(w;x,\alpha)\right|_{w=w_+}=-2\frac{(1+\sqrt{\alpha})^4}{\sqrt{\alpha}}\;.
\end{align}
Therefore, the steepest descent directions at the point $w_{+}$ are $\{2\pi/3,4\pi/3,0\}$. One can locally deform the
contour, introducing the scalar $t\in(-\infty,+\infty)$, approaching and leaving $w_+$ with $w(t)=e^{i4\pi/3}t+w_{+}$ and $w(t)=e^{i2\pi/3}t+w_{+}$ respectively, yielding
\begin{align}
\frac{\mathcal{Z}_{\beta,1}(\lambda)}{\lambda^N}&=\frac{e^{N{\cal L}_1(w_+;x,\alpha)}}{2i \pi w_+}\Bigg(e^{i\frac{2\pi}{3}}\int_{0}^{+\infty}dt e^{-N\frac{2(1+\sqrt{\alpha})^4}{3!\sqrt{\alpha}}t^3}+e^{i\frac{4\pi}{3}}\int_{+\infty}^{0}dt e^{-N\frac{2(1+\sqrt{\alpha})^4}{3!\sqrt{\alpha}}t^3}\Bigg)\\
&=\frac{\Gamma(1/3)}{2\pi\,3^{1/6}\big(N\frac{(1+\sqrt{\alpha})}{\sqrt{\alpha}}\big)^{1/3}}e^{N{\cal L}_1(w_+;x,\alpha)}\nn
\end{align}

\noindent
2) For $x=x_-(\alpha)$, $w_{-}=(1-\sqrt{\alpha})^{-1}$, such that
\begin{align}
\left.\partial_{w}^2{\cal L}_1(w;x,\alpha)\right|_{w=w_-}=&0\;,\;\;\left.\partial_{w}^3{\cal L}_1(w;x,\alpha)\right|_{w=w_-}=2\frac{(1-\sqrt{\alpha})^4}{\sqrt{\alpha}}\;.
\end{align}
The steepest descent direction at $w_{-}$ are $\{\pi/3,\pi,5\pi/3\}$. The contour is deformed by parametrizing the path, approaching and leaving $w_{-}$ with $w(t)=e^{i1/3\pi}t+w_{-}$ and $w(t)=e^{i5/3\pi}t+w_{-}$ respectively, yielding
\begin{align}
\frac{\mathcal{Z}_{\beta,1}(\lambda)}{\lambda^N}&=\frac{e^{N{\cal L}_1(w_-;x,\alpha)}}{2i \pi w_-}\Bigg(e^{i\frac{\pi}{3}}\int_{0}^{+\infty}dt e^{-N\frac{2(1-\sqrt{\alpha})^4}{3!\sqrt{\alpha}}t^3}+e^{i\frac{5\pi}{3}}\int_{+\infty}^{0}dt e^{-N\frac{2(1-\sqrt{\alpha})^4}{3!\sqrt{\alpha}}t^3}\Bigg)\\
&=\frac{\Gamma(1/3)(-1)^N}{2\pi\,3^{1/6}\big(N\frac{(\sqrt{\alpha}-1)}{\sqrt{\alpha}}\big)^{1/3}}e^{N{\cal L}_1(w_-;x,\alpha)}\nn
\end{align}
In each case, one obtains that the term 
\begin{align}
\Xi_{1}(x=x_{\pm}(\alpha);\alpha)=&\lim_{N\rightarrow+\infty}\frac{1}{N}\ln\left|\frac{\mathcal{Z}_{\beta,1}(\lambda)}{\lambda^N}\right|={\cal L}_1(w_\pm;x_{\pm}(\alpha),\alpha)\;,   
\end{align}
where $x=\lambda/\moy{T}$.

\section{Finite $N$ expression of ${\cal Z}_{\beta,2}$} \label{app_Z_2_b_N_fin}

\noindent Let us first consider $\beta=2$. In this case ${\Lambda}=\diag(\lambda_1,\lambda_2)$ and we introduce 4 variables $\psi_{ij}$ ,with $i\le j$ and $i,j=1,2$, and $\phi$ such that $0\le \psi_{ij}\le 2\pi$ and $0\le \phi \le\pi/2$. Therefore a possible parametrization of $ U$ is the following \cite{Dita82}: 
\be
 U = \begin{bmatrix}
   e^{i(\psi_{11}+\psi_{12})}\cos\phi &  e^{i\psi_{12}}\sin\phi \\
 -e^{i(\psi_{11}+\psi_{22})}\sin\phi  &   e^{i\psi_{22}}\cos\phi
 \end{bmatrix}
\ee
The Eucledean line element $(\mathrm{d}s)^2=\sum_{i,j}|\mathrm{d} U_{ij}|^2$ reads in this parametrisation
\begin{align}\label{dssquared}
    (\mathrm{d}s)^2=&(d\psi_{11})^2+(d\psi_{12})^2+(d\psi_{22})^2+2(d\phi)^2\\
&+2(\cos\phi)^2(d\psi_{11})(d\psi_{12})+2(\sin\phi)^2(d\psi_{11})(d\psi_{22})\;.\nn
\end{align}
The associated Jacobian is given by $\sqrt{2\det{M}}$
where:
\be
{M}=\begin{bmatrix}
  1& (\cos\phi)^2 & (\sin\phi)^2\\
  (\cos\phi)^2& 1& 0\\
  (\sin\phi)^2& 0&1\\
 \end{bmatrix}
\ee
The normalised Haar measure is obtained calculating $\int\sqrt{\det{M}}d\psi_{11}d\psi_{12}d\psi_{22}d\phi$ and is given by $d U=\frac{1}{4\pi^3}\cos\phi\sin\phi d\psi_{11}d\psi_{12}d\psi_{22}d\phi$. With the further assumption $\moy{T}=0$, from what stated above follows 
\begin{align}
    &\det U=e^{i(\psi_{11}+\psi_{12}+\psi_{22})}\;,\\
    &\Tr{\Lambda} U=\cos\phi(e^{i(\psi_{11}+\psi_{12})}\lambda_1+e^{i(\psi_{22})}\lambda_2)\;,\\
    &\moy{\det(\mathbb{I}_2-\frac{T_a}{N} U)}=1+\frac{\moy{T^2}}{N^2}e^{i(\psi_{11}+\psi_{12}+\psi_{22})}\;.
\end{align}
The denominator of Eq.(\ref{eq1U}) reads
\begin{align}\label{b21}
&\frac{1}{4\pi^3}\int_{0}^{2\pi}d\psi_{11}\int_{0}^{2\pi}d\psi_{12}\int_{0}^{2\pi}d\psi_{22}d\phi\cos\phi\sin\phi e^{\cos\phi(e^{i(\psi_{11}+\psi_{12})}+e^{i\psi_{22}})}e^{-iN(\psi_{11}+\psi_{12}+\psi_{22})}\\
&=\frac{2}{(N!)^2}\int_{0}^{\frac{\pi}{2}}d\phi (\cos\phi)^{2N+1}\sin\phi=\frac{1}{(N!)^2}\frac{1}{N+1}\;,\nn
\end{align}
while the numerator reads
\begin{align}\label{b22}
&\frac{1}{4\pi^3}\int_{0}^{2\pi}d\psi_{11}\int_{0}^{2\pi}d\psi_{12}\int_{0}^{2\pi}d\psi_{22}d\phi\cos\phi\sin\phi \\
&\times e^{\cos\phi(e^{i(\psi_{11}+\psi_{12})}\lambda_1+e^{i(\psi_{22})}\lambda_2)}e^{-iN(\psi_{11}+\psi_{12}+\psi_{22})}\Big(1+\frac{\moy{T^2}}{N^2}e^{i(\psi_{11}+\psi_{12}+\psi_{22})}\Big)^{M}\nn\\
&=\sum_{k=0}^{N}\frac{2}{k!(N-k!)!}\frac{(\lambda_1\lambda_2)^k}{(N-k)!}\frac{\alpha N!}{(\alpha N-k)!}\Big(\frac{\moy{T^2}}{N^2}\Big)^k\frac{1}{2-2k+2N}\,.\nn
\end{align}
Therefore, for $\beta=2$, Eq.(\ref{eq3}) is given by the ratio of Eq.(\ref{b22}) and Eq.(\ref{b21}).

\noindent
For $\beta=1$, we impose $\moy{T^2}=1$ to simplify the computations. The general case is retrieved by multiplying ${\cal Z}_{2,\beta=1}({\bf \lambda})$ by $\moy{T^2}^N$ and rescaling $\lambda_{1,2}\rightarrow\lambda_{1,2}/\sqrt{\moy{T^2}}$. We rename ${\Lambda}=\diag(\lambda_1,\lambda_2,\lambda_1,\lambda_2)$ while now the matrix $U$ is parametrized with a $2\times2$ block matrix. Lastly, we introduce $6$ variables: $h_1,h_2\in(0,1)$ and $\phi_{11},\phi_{21},\phi_{22},\phi_{14}\in(0,2\pi)$.
Therefore (see \cite{Dita82}),
\be
 U = \begin{bmatrix}
    U_{11} &   U_{12}\\\
 - U_{12}e^{i(\phi_{12}+\phi_{21}-2\phi_{14})} &   U_{11}^t
 \end{bmatrix}
\ee
with
\be
 U_{11}= \begin{bmatrix}
   e^{i\phi_{11}}h_1 &  e^{i\phi_{22}}h_2\sqrt{1-h_1^2}\\\
 e^{i\phi_{21}}h_2\sqrt{1-h_1^2} &  -e^{-i(\phi_{11}-\phi_{21}-\phi_{22})}h_1
 \end{bmatrix}
\ee
and
\be
 U_{12}=e^{i\phi_{14}}\sqrt{1-h_1^2}\sqrt{1-h_2^2} \begin{bmatrix}
   0 &  -1\\
1 &  0
 \end{bmatrix}
\ee
Similarly to Eq.(\ref{dssquared}), in order to retrieve the normalized Haar measure, the Eucledean line element is $(\mathrm{d}s)^2=\sum_{i,j}|\mathrm{d} U_{ij}|^2=2\sum_{i,j}|\mathrm{d} U_{11;i,j}|^2+\sum_{i,j}|\mathrm{d} U_{12;i,j}|^2+\sum_{i,j}|\mathrm{d} U_{21;i,j}|^2$. In details we have that:
\begin{align}
(ds)^2=&2((dh_1)^2+h_1^2(d\phi_{11})^2+2(1-h_1^2)(dh_2)^2-4dh(dh_1)(dh_2)\\
&+\frac{2(dh_1)^2}{(1-h_1^2)}(dh_1)^2+h_2^2(1-h_1^2)(d\phi_{22})^2+h_2^2(1-h_1^2)(d\phi_{21})^2+(dh_1)^2\nn\\
&+h_1^2(d\phi_{22}+d\phi_{21}-d\phi_{11})^2+\frac{2(h_2(1-h_1^2)(dh_2)+h_1(1-h_2^2)(dh_1))^2}{(1-h_2^2)(1-h_1^2)}\nn\\
&+(1-h_2^2)(1-h_1^2)(d\phi_{14})^2+(1-h_2^2)(1-h_1^2)(d\phi_{22}+d\phi_{21}-d\phi_{14})^2)\nn
\end{align}
To simplify the calculation, we can re-arrange the differential of the 6 variables such that the Jacobian is given by the product of the square roots of the matrices ${A}_1$ and ${A}_2$. The latter are:
\be
{A}_1=4\begin{bmatrix}
   1+\frac{h_1^2(1-h_2^2)+h_2^2h^2}{(1-h_1^2)} &  0\\
0 &  (1-h_1^2)+\frac{h_2^2(1-h_1^2)}{(1-h_2^2)}
 \end{bmatrix}
\ee
and
\be
{A}_2=
\begin{bmatrix}
   4h^2 & -2h^2& -2h^2&  0\\
 -2h^2 & 2&2-2h_2^2+2d^2h^2&-2(1-h_2^2)(1-h_1^2)\\
  -2h^2 & 2-2h_2^2+2h_2^2h^2& 2 &  -2(1-h_2^2)(1-h_1^2)\\
   0 & -2(1-h_2^2)(1-h_1^2)& -2(1-h_2^2)(1-h_1^2)& 4(1-h_2^2-h_1^2+h_2^2h^2)\\
 \end{bmatrix}.
\ee
We obtain the Haar measure by calculating 
\begin{align}
&\int \sqrt{\det({A}_1)\det({A}_2)} dh_1 dh_2 d\phi_{11}d\phi_{21}d\phi_{22}d\phi_{14}\;,\\
&\det({A}_1)=\frac{16}{1-h_2^2}\;,\\
&\det({A}_2)=64h_2^2(1-h_2^2)h_1^2(1-h_1^2)^2\;.
\end{align}
Integrating out the 6 linear and angular variables, one obtains
\be
d U=\frac{1}{2\pi^4}h_1 h_2(1-h_1^2)(\mathrm{d} h_1\mathrm{d} h_2 \mathrm{d}\phi_{11}\mathrm{d}\phi_{21}\mathrm{d}\phi_{22}\mathrm{d}\phi_{14})\;.
\ee 
The denominator appearing in Eq.(\ref{eq1U}) is
 \begin{align}
   & \frac{1}{2\pi^3}\int_{0}^{1}\mathrm{d}h_1\int_{0}^{2\pi}\mathrm{d}\phi_{11}\int_{0}^{2\pi}\mathrm{d}\phi_{21}\int_{0}^{2\pi}\mathrm{d}\phi_{22}e^{he^{-i\phi_{11}}(e^{2i\phi_{11}}-e^{i(\phi_{21}+\phi_{22})})}e^{-iN(\phi_{21}+\phi_{22})}=\\ 
   &\frac{4}{(N!)^2}\int_{0}^{1}\mathrm{d}h_1 h_1^{2n+1}(1-h_1^2)=\frac{2}{(N!)^2}\frac{1}{(2+3N+N^2)}\;,
 \end{align}
whereas the numerator in Eq.(\ref{eq1U}) reads
 \begin{align}\label{NumeratoreB14}
&\frac{1}{2\pi^3}\int_0^{1}\mathrm{d}h_1\int_{0}^{2\pi}\mathrm{d}\phi_{11}\int_{0}^{2\pi}\mathrm{d}\phi_{21}\int_{0}^{2\pi}\mathrm{d}\phi_{22}h_1(1-h_1^2)\\
&\times e^{-iN(\phi_{22}+\phi_{21})} e^{h_1\Big(\lambda_1e^{i\phi_{11}}-\lambda_2e^{i(\phi_{22}+\phi_{21}-\phi_{11})}\Big)}\Big(1-\frac{e^{i(\phi_{22}+\phi_{21})}}{N^2}\Big)^{M}\,.\nn
 \end{align}
The angular variables $\phi_{22}$ and $\phi_{21}$ in the integral above can be integrated out with the auxiliary variable $\phi=\phi_{22}+\phi_{21}$. The integral appearing in Eq(\ref{NumeratoreB14}) is simplified to
\be
\sum_{k=0}^N\frac{2(\alpha N!)\,N^{2(\alpha N-k)}}{\pi N^{2\alpha N}\,k!(N-k)!(\alpha N-k)!}\int_{0}^1\mathrm{d}h_1 h_1(1-h_1^2)(h_1\lambda_2)^{N-k}\int_{0}^{2\pi}\mathrm{d}\phi_{11}e^{h_1\lambda_1 e^{i\phi_{11}}-i(N-k)\phi_{11}}\,.     
\ee
Working out the remaining integration, the expression above becomes 
\begin{align}
\sum_{k=0}^N\frac{2(\alpha N!)(\lambda_1\lambda_2)^N}{(k!(\lambda_1\lambda_2N^2)^k)((N-k)!)^2 (\alpha N-k)(2+3(N-k)+(N-k)^2)!}
\end{align}
and Eq(\ref{eq3}) is retrieved with the introduction of the hypergeometric function.

\section{Laplace approximation for ${\cal Z}_{\beta,2}$} \label{app_LD_Z_b_2}
We consider the case of centred random variables, i.e. $\moy{T}_i=0$. In order to obtain the asymptotics of ${\cal Z}_{\beta,2}$, it is convenient to start from Eq.(\ref{eq3}) and to introduce $x=\lambda_1\lambda_2/\moy{T^2}$. Expanding the terms within the  hypergeometric function, we notice
\begin{align}
\frac{\mathcal{Z}_{\beta,2}(\bm{\lambda})}{(\lambda_1\lambda_2)^N}=&\sum_{k=0}^N\frac{\beta}{2}\frac{ (N-k+1)_k  \Gamma (\alpha N+1) \Gamma (N+\frac{2 }{\beta}+1)}{\left(\frac{2}{\beta}+1\right)_k \Gamma ((\alpha -1)N+1) ((\alpha -1)N+1)_k}\frac{\left(N^2 x\right)^{k-N}}{k!}\\
&=\sum_{k=0}^N a_N(x;k)\;,\nn
\end{align}
where $(b)_k=\frac{\Gamma(b+k)}{\Gamma(b)}$ is the Pochhammer symbol. 

Let us first consider the case where $x>0$. In that case, we can safely replace in the large $N$ limit the summation above with an integral, by rescaling $k=N\eta$ with $\eta\in[0,1]$. Therefore, for $N$ sufficiently large, one observes
\begin{align}
a_N(x;N\eta)\approx &\frac{\beta\sqrt{\frac{2\alpha}{\pi }} \Gamma \left(\frac{2}{\beta}+1\right) \eta^{-\frac{2}{\beta}-1}}{4  \sqrt{N} \sqrt{1-\eta} \sqrt{(\alpha +\eta-1)}} e^{N{\cal L}_2(\eta;x,\alpha)}\;,\\
{\cal L}_2(\eta;x,\alpha)=&\alpha\ln(\alpha)-(\alpha+\eta-1)\ln (\alpha+\eta-1)\\
&+\eta\ln (x(1-\eta ))+2\eta-2\eta\ln(\eta)-\ln (1-\eta)-2-\ln u\nn\;.
\end{align}
The maximum of $\mathcal{L}_2$ is located in $\eta_*\in[0,1]$. The latter is the solution of the equation
\begin{align}
\left.\partial_{\eta}\mathcal{L}_2\right|_{\eta=\eta_*}=0\;,\;\;\Rightarrow\;\;\ln (x-\eta_* x)-\ln(\alpha +\eta_*-1)-2 \ln(\eta_*)=0\;.
\end{align}
Hence, we need to retrieve the roots of $p(\eta)=\eta^2(\alpha-1+\eta)-x(1-\eta)$. Firstly, one  observes $\lim_{\eta\rightarrow\pm\infty}p(\eta)=\pm\infty$ and $p(0)=-x$. For $\eta>0$, $p(\eta)$ is strictly increasing since $\eta^2+2\eta(\alpha+\eta-1)+x>0$. For $x\rightarrow 0^+$, the zero of $p$ tends to $0$. For $x\rightarrow+\infty$ it tends to $1$. With the help of Cardano's formula, the root is
\begin{align}
&\eta_*=\frac{1-\alpha }{3}-\frac{\sqrt[3]{2} \Theta}{3 \sqrt[3]{\Delta +\sqrt{\Delta ^2+4 \Theta^3}}}+\frac{\sqrt[3]{\Delta+\sqrt{\Delta ^2+4 \Theta^3}}}{3 \sqrt[3]{2}},\nn\\
&\Delta=9x(2+\alpha)-2(\alpha-1)^3\;,\;\;\Theta=3 x-(\alpha -1)^2\;.
\end{align}
One can check that this solution $\eta_*\in(0,1)$ and in this interval corresponds to a maximum as 
\be
\left.\frac{d^2\mathcal{L}}{d\eta^2}(\eta)\right|_{\eta=\eta_*}=-\frac{2}{\eta_*}+\frac{\alpha}{(\eta_*-1)(\eta_*+\alpha-1)}<0\;.
\ee
Finally, one obtains that in the regime $N\gg 1$, 
\begin{align}
\frac{\mathcal{Z}_{\beta,2}(\bm{\lambda})}{(\lambda_1\lambda_2)^N}\approx &\frac{\beta\sqrt{\frac{2}{\pi }} \Gamma \left(\frac{2}{\beta}+1\right) {\eta^{*}}^{-\frac{2}{\beta}-1} \sqrt{\alpha  N}}{4\sqrt{1-\eta_*} \sqrt{(\alpha +\eta_*-1)}} e^{N\mathcal{L}(\eta_*)}\label{res_x_pos}\int_{-\infty}^{+\infty} d\eta e^{\frac{N}{2}\frac{d^2\mathcal{L}}{d\eta^2}(\eta_*)(\eta-\eta_*)^2}\;.
\end{align}
Let us now consider the more involved case where $x<0$. In that case, one needs to consider separately the terms for even and odd values of $k$ by re-writing the expression as 
\begin{align}
\frac{\mathcal{Z}_{\beta,2}(\bm{\lambda})}{(\lambda_1\lambda_2)^N}=&\sum_{k=0}^{N/2}\frac{(N-2k+1)_{2k+1}}{(1+\frac{2}{\beta})_{2k}((\alpha-1)N+1)_{2k}}\frac{(N^2 |x|)^{2k-N}}{(2k)!}\\
&-\sum_{k=0}^{N/2-1}\frac{(N-2k)_{2k+1}}{(1+\frac{2}{\beta})_{2k+1}((\alpha-1)N+1)_{2k+1}}\frac{(N^2 |x|)^{2k+1-N}}{(2k+1)!}\;. \nn 
\end{align}
We may now in each term replace the sum over $k$ with an integral, over respectively $\eta_{\rm e}=2k/N$ and $\eta_{\rm o}=(2k+1)/N$ with $\eta_{\rm o},\eta_{\rm e}\in[0,1]$. The saddle point is the same as for $x>0$ but one now needs to substract the contribution from even and odd values of $k$. This is equivalent to doing a Taylor expansion in the parameter $\eta_*$ of the expression in Eq. \eqref{res_x_pos}. The result at exponential order in $N$ is the same but the finite $N$ corrections are thus different.


\begin{thebibliography}{}
\bibitem{footnote}
One can also consider  taking $K$ with  quaternion independent normally-distributed elements to generate the ensemble with Dyson index $\beta=4$.
\end{thebibliography}
\end{document}